\documentclass[nofootinbib,prd,superscriptaddress,twocolumn]{revtex4}

\usepackage{amsmath}
\usepackage{amsfonts}
\usepackage{amssymb}
\usepackage{graphicx}
\usepackage[titletoc]{appendix}
\usepackage{color}
\usepackage{hyperref}
\usepackage{cleveref}
\usepackage[rightcaption]{sidecap}
\usepackage{comment}
\usepackage{soul}
\usepackage{cancel}

\usepackage{dcolumn}

\usepackage{longtable}
\usepackage{floatrow}
\usepackage{array}
\usepackage{ctable}
\usepackage{multirow}
\usepackage{siunitx}
\usepackage{tabularx}
\usepackage{booktabs}
\usepackage{supertabular}

\graphicspath{{Graphics/}}

\def\be{\begin{equation}}
\def\ee{\end{equation}}
\def\bea{\begin{eqnarray}}
\def\eea{\end{eqnarray}}

\def\prl{Phys. Rev. Lett.}

\def\prd{Phys. Rev. D}
\def\mnras{MNRAS}

\def\apj{ApJ}
\def\apjl{ApJ Lett.}
\def\apjs{ApJ Suppl. Ser.}

\def\aap{A\&A}

\def\araa{Annual Rev. of Astron. Astrophys.}

\def\physrep{Phys. Rep.}
\def\jcap{JCAP}

\definecolor{vividviolet}{rgb}{0.62, 0.0, 1.0}
\definecolor{amaranth}{rgb}{0.9, 0.17, 0.31}
\definecolor{palatinateblue}{rgb}{0.15, 0.23, 0.89}
\definecolor{brightpink}{rgb}{1.0, 0.0, 0.5}
\definecolor{cornflowerblue}{rgb}{0.39, 0.58, 0.93}
\definecolor{deepcarminepink}{rgb}{0.94, 0.19, 0.22}
\definecolor{radicalred}{rgb}{1.0, 0.21, 0.37}

\hypersetup{ linktoc=all,
    colorlinks, linkcolor={palatinateblue},
    citecolor={brightpink}, urlcolor={amaranth}
}

\begin{document}

\title{Cosmic distance duality after DESI 2024 data release and dark energy evolution}

\author{Anna Chiara Alfano}
\email{a.alfano@ssmeridionale.it}
\affiliation{Scuola Superiore Meridionale, Largo S. Marcellino 10, 80138 Napoli, Italy.}
\affiliation{Istituto Nazionale di Fisica Nucleare (INFN), Sezione di Napoli Complesso Universitario Monte S. Angelo, Via Cinthia 9 Edificio G, 80138 Napoli, Italy.}

\author{Orlando Luongo}
\email{orlando.luongo@unicam.it}
\affiliation{Universit\`a di Camerino, Divisione di Fisica, Via Madonna delle carceri 9, 62032 Camerino, Italy.}
\affiliation{INFN, Sezione di Perugia, Perugia, 06123, Italy.}
\affiliation{INAF, Osservatorio Astronomico di Brera, Milano, Italy.}
\affiliation{Al-Farabi Kazakh National University, Al-Farabi av. 71, 050040 Almaty, Kazakhstan.}

\begin{abstract}
The cosmic distance duality  relates the angular-diameter and luminosity distances and its possible violation may puzzle the standard cosmological model. This appears particularly interesting in view of the recent results found by the DESI Collaboration, suggesting that a dynamical dark energy scenario seems to be favored than a genuine cosmological constant. Accordingly, we take into account possible violations by considering four different parameterizations, namely: a Taylor expansion around $z\simeq 0$, a slightly-departing logarithmic correction, a (1;2) Pad\'e rational series to heal the convergence problem and a Chebyshev  polynomial expansion, reducing \emph{de facto} the systematic errors associated with the analysis. We test each of them  in a model-independent (-dependent) way, by working out Monte-Carlo Markov chain analyses, employing the B\'ezier interpolation of the Hubble rate $H(z)$ for the model-independent approach while assuming the flat (non-flat) $\Lambda$CDM and  $\omega_0\omega_1$CDM models, motivating the latter paradigm in view of the DESI findings. Subsequently, we explore two analyses, employing observational Hubble data, galaxy clusters from the Sunyaev–Zeldovich effect and type Ia supernovae, investigating the impact of the DESI data catalog, first including then excluding the entire data set. Afterwards, we adopt statistical model selection criteria to assess the statistically favored cosmological model. Our results suggest \emph{no violation} of the cosmic distance duality. While a slight spatial curvature cannot be entirely excluded, the preferred cosmological model remains the flat $\Lambda$CDM background, even when incorporating DESI data. Finally, concerning the Hubble tension, our findings match the Riess estimates, as BAO data points are excluded.
\end{abstract}

\maketitle
\tableofcontents

\section{Introduction}

In modern cosmology, the number of data at our disposal is drastically increasing, enabling us to achieve more stringent constraints on the cosmological parameters, with the aim of clarifying the dark energy nature\footnote{We are currently experiencing an era of \emph{precision cosmology} \cite{Bamba:2012cp}, where, quite unprecedentedly, a \emph{plethora} of data is revealing new effects,  allowing a direct comparison with the various theoretical models.}. Remarkably, the recent outcomes from the \citet{2024arXiv240403002D, 2024arXiv241112022D} seem to point towards the need of considering an \emph{evolving dark energy} instead of a pure cosmological constant, $\Lambda$ \cite{weinberg1989cosmological, 1992ARA&A..30..499C, 2001LRR.....4....1C, 2003PhR...380..235P,Luongo:2018lgy,Luongo:2021nqh}.

Nevertheless, even though the $\Lambda$CDM paradigm appears  statistically favored, it suffers from theoretical and observational issues. Specifically, a fine-tuning problem, i.e., a discrepancy of 121 orders of magnitudes between the theoretical and observational values of $\Lambda$ and a coincidence problem, i.e., the strange comparability between the density of the cosmological constant and of matter at late-times \cite{2006IJMPD..15.1753C}. Further, there also appear evident tensions between low and high-redshift values of the Hubble constant $H_0$ and the amplitude of clustering $S_8$ \cite{2022JHEAp..34...49A, 2021APh...13102605D, 2021APh...13102604D}.

In this puzzle,  several dark energy scenarios have been proposed across recent years,  trying to explain possible  departures from the concordance paradigm\footnote{For the sake of clearness, albeit the DESI collaboration emphasizes the need of an evolving scalar field dark energy, severe criticisms against these findings have been raised, see e.g. Refs. \cite{2024A&A...690A..40L, 2024arXiv240408633C, 2024arXiv241212905C, 2024JCAP...12..055A, 2024JCAP...09..062D}.} \cite{2012IJMPD..2130002Y, 2024arXiv240412068C, 2024A&A...690A..40L, 2024arXiv241104901L, 2024JCAP...12..055A, 2024arXiv241104878A, 2024PhRvD.110h3528W, 2024arXiv240917019W, 2024arXiv240408633C, 2024arXiv240917074G, 2024JCAP...10..035G, 2024PhRvD.109l1305T}.

To check the consistency of the $\Lambda$CDM model, one can seek possible violation of the cosmic distance duality (CDD), emphasizing insights on the dark energy nature.

The CDD relation states that the ratio between the luminosity and angular-diameter distances is equal to the inverse square of the scale factor $a(t)$. If a deviation from such a relation occurs, then the reasons could be,

\begin{itemize}
    \item[i.] either systematics associated with the measurements \cite{2004PhRvD..69j1305B},
    \item[ii.] or direct deviations from the concordance paradigm \cite{ellis2007definition}.
\end{itemize}

In this respect, the possibility of a violation of the CDD relation has already been explored, adopting model-dependent or -independent approaches\footnote{For completeness, so far no model-independent technique has raised robust evidences against the standard paradigm, even invoking new cosmographic parameters, such as transition and equivalence times \cite{Capozziello:2021xjw}, or extensions of Taylor series at higher redshifts \cite{Luongo:2020aqw,Luongo:2022bju} or alternatives to standard analyses, see e.g. Refs. \cite{Luongo:2024pcp,Luongo:2024zhc}. The CDD violation can therefore offer an alternative to standard avenues.}, with no appreciable evidence \cite{2004PhRvD..70h3533U, 2004PhRvD..69j1305B, 2012JCAP...06..022H, 2012ApJ...745...98M, Jesus:2024nrl, 2017JCAP...09..039H, 2011A&A...528L..14H, 2021MNRAS.504.3938M}. However, the  DESI results, showing a more favored slightly evolving dark energy fluid, not under the form of a cosmological constant, reopens to check evidences in favor of a CDD violation.

Motivated by the above considerations, we here consider that a possible violation of the CDD relation acquires the formal structure,

\begin{equation}\label{cddviol}
    \frac{d_L(z)}{d_A(z)(1+z)^2} = \eta(z),
\end{equation}
that violates the CDD relation as  $\eta(z)\neq 1$. Hereafter, we thus employ \emph{four different $\eta(z)$ parameterizations}.

The first is the standard  Taylor expansion around $z\simeq 0$ to map regions at $z\leq 1$, the second is represented by a logarithmic  correction, obtained from a polynomial ansatz over $\eta$, also used to investigate possible deviations from the cosmic transparency \cite{2009JCAP...06..012A}, while the third and fourth approaches represent robust alternatives to the standard Taylor expansions, motivated by the issue of convergence jeopardizing cosmographic analyses at background, see e.g. Refs. \cite{Aviles:2012ay, Aviles:2014rma, Cattoen:2007sk, Cattoen:2007id}.  Precisely, we employ the (1;2) Pad\'e series, represented by a rational expansion in terms of two orders and well-matching as the redshift increases and the fourth provided by the use of Chebyshev  polynomials that exhibit the advantage of significantly reducing the systematics  \cite{Gruber:2013wua,2018MNRAS.476.3924C}. Hence, as stated above, the comparison with the here-adopted alternative expansions, i.e., Pad\'e and Chebyshev, appears useful not only to heal the convergence problem but also to reduce the uncertainties on estimating the key  parameters under exam. Afterwards, we adopt and compare two distinct methodologies considering a model-independent approach employing  the B\'ezier interpolation of the Hubble rate $H(z)$ up to the second order \cite{amati2019addressing} and, then, a model-dependent approach where we consider both flat and non-flat $\Lambda$CDM and $\omega_0\omega_1$CDM models \cite{2020A&A...641A...6P, 2001IJMPD..10..213C, 2003PhRvL..90i1301L}. For both the aforementioned approaches,  we consider two independent cosmological fits employing a Monte-Carlo Markov chain (MCMC) simulation adopting the Metropolis-Hastings algorithm to get constraints on the parameters of our interest \cite{1953JChPh..21.1087M, 1970Bimka..57...97H}. The data sets, adopted in our analyses, consist in the updated sample of the observational Hubble data (OHD), galaxy clusters from which the angular-diameter distances are determined, the Pantheon catalog of type Ia supernovae (SNeIa) and the recently baryonic acoustic oscillation (BAO) data points from the DESI first release.

More precisely, to explore the new DESI data impact, in the first analysis, labeled \emph{Analysis 1}, we involve data points argued from the  \citet{2024arXiv240403002D} while in the second, labeled \emph{Analysis 2}, we exclude them, comparing our constraints between them. Afterwards, we also assess the values inferred from both our analyses with the ones got from the \citet{2020A&A...641A...6P}, for both spatially flat and non-flat scenarios.

Moreover,  our findings on the (reduced) Hubble constant,  $h_0$, are reanalyzed in terms of the Hubble tension\footnote{Two different measurements, providing $h_0 = 0.674\pm 0.005$ from the Planck mission \cite{2020A&A...641A...6P} and $h_0 = 0.730\pm 0.010$, derived from SNeIa \cite{Riess:2021jrx}, appear strongly in tension, likely indicating new physics \cite{2021CQGra..38o3001D, 2017PhRvD..96d3503D, 2020PhRvD.102f3527N, 2023PDU....3901163H, 2023PhRvD.108d3513H}.} and noticing that our bounds appear more compatible with the Riess results, albeit not fully excluding the Planck constraints up to $2$-$\sigma$.

Last but not least, adopting model selection criteria, i.e., the Akaike information criterion (AIC) and the deviance information criterion (DIC) \cite{akaike1998information, spiegelhalter2002bayesian,  2006PhRvD..74b3503K}, we also check the preferred cosmological model, finding that the spatially flat $\Lambda$CDM model remains the favored paradigm to the detriment of the $\omega_0\omega_1$CDM scenario. Additionally and quite remarkably, we do not stress evidence favoring the CDD violation, albeit it is worth stressing that a non-zero spatial curvature is not excluded from both the model-independent (-dependent) treatments.

The paper is organized as follows. In Sect. \ref{CDD} we introduce the CDD relation and the four parameterizations we are going to use when the relation is violated. Then, Sect. \ref{cosmicdata} deals with the description of the model-independent technique and cosmological models we are going to use together with cosmic data sets adopted in our computations. In Sects. \ref{results}-\ref{mselec} we expose the results inferred from our analyses and the ones from the model selection criteria, respectively. Finally, Sect. \ref{conc} discusses conclusions and perspectives.

\section{The cosmic distance duality relation}\label{CDD}

Etherington's reciprocity law \cite{etherington1933lx}, also known as CDD relation, relates the luminosity, $d_L(z)$, and the angular-diameter, $d_A(z)$,  distances through the relation,

\begin{equation}\label{cdd}
    \frac{d_L}{d_A} = a^{-2}.
\end{equation}

In precision cosmology, Eq. \eqref{cdd} is true for \emph{any cosmological model} within a cosmic background in which the cosmological principle holds and, in particular, where photons travel along null geodesics and their number is conserved \cite{2020A&A...644A..80M}.

The interest in Eq. \eqref{cdd} lies on the fact that seeking deviations from it implies either possible systematic errors in observations or  departures from the standard $\Lambda$CDM model \cite{2004PhRvD..69j1305B, Jesus:2024nrl}. Departing from Eq. \eqref{cdd} and setting $a=(1+z)^{-1}$ turn into the CDD relation , written in the form \eqref{cddviol}.

However, the CDD violation appears even more general than violating Eq. \eqref{cdd} only, as commonly discussed in the literature, see e.g. Refs. \cite{2017JCAP...09..039H, 2012JCAP...06..022H, 2011A&A...528L..14H, Jesus:2024nrl, Wang:2024rxm}. Indeed, it is immediately evident that distance definitions in cosmology are quite common, and any CDD violation may turn into deviations in either $d_L$ or  $d_A$ or even on both simultaneously.

Consequently, one cannot \emph{a priori} determine whether a possible violation originates from one of the two distances or affects both. In addition, such a violation would impact other distance definitions, including the photon flux distance, $d_F$, the photon count distance, $d_P$, and the deceleration distance, $d_Q$. In fact, baptizing,
\begin{equation}\label{r00}
r_0 = \int_{t}^{t_0}{\frac{dt'}{a(t')}}\,,
\end{equation}
as comoving distance, used to define $d_L$ by $d_L  =  a_0 r_0 (1+z) = r_0 \cdot \frac{1}{a(t)}$ or $d_A$ by $d_A  =  \frac{d_L}{(1+z)^2} = r_0\cdot a(t)$, we can easily define
\begin{subequations}\label{serie2}
\begin{align}
    d_F & =  \frac{d_L}{(1+z)^{1/2}} = r_0 \cdot \frac{1}{\sqrt{a(t)}}\,,\\
    d_P & =  \frac{d_L}{(1+z)} = r_0\,,\\
    d_Q & =  \frac{d_L}{(1+z)^{3/2}} = r_0 \cdot \sqrt{a(t)}\,.
    \label{eq:dist}
\end{align}
\end{subequations}

These above alternative distances are less frequently debated in the literature but they turn out to be physically relevant. Precisely,

\begin{itemize}
\item[-] the photon flux distance, $d_F$, is derived from the photon flux rather than the energy flux at the detector, making it experimentally more accessible;
\item[-] the photon count distance, $d_P$, based on the total number of photons arriving at the detector, as opposed to the photon arrival rate;
\item[-] the deceleration distance, $d_Q$, providing a simple and practical dependence on the deceleration parameter $q_0$ ~\cite{Cattoen:2008th, Aviles:2012ay}.
\end{itemize}

Since a fully-determined distance expansion requires computing the comoving distance traveled by a photon from a light source at $r=r_0$ and to an observer at $r=0$ - a CDD violation would have direct implications for flux measurements, photon counts, and related observable quantities.

Its significance lies in the modifications of the physical definitions within Eq.~\eqref{r00}.

Since determining \emph{any distance expansion} requires computing $r_0$, a violation of the CDD would directly affect it, and consequently impact flux measurements, photon counts, and related quantities. Thus, its significance lies \emph{not only} in the modification of the physical definition in Eq. \eqref{r00} but also, notably, in the implications for the physics of \emph{any cosmic distances} used in a given context\footnote{This turns out to be evident checking cosmographic expansions on background of $r_0$, see e.g. Ref. \cite{Aviles:2012ay}.}, provided that its definition depends on $r_0$.

Accordingly, to test Eq. \eqref{cddviol} different parameterizations of $\eta(z)$ have been proposed \cite{2017JCAP...09..039H, 2014JPhCS.484a2035J, Jesus:2024nrl, 2020PhRvD.102f3513D, 2013PhRvD..87j3530E, 2021PhRvD.103j3513A, 2021JCAP...06..052B, 2021MNRAS.502.3500Q, 2012JCAP...06..022H, 2011A&A...528L..14H, 2019PhRvD..99h3523F}. Accordingly, there is a vast literature that tries to check departures through the use of Eq. \eqref{cddviol}, albeit robust evidences for $\eta\neq1$ have not been found so far.

We here choose four different frameworks to check the CDD, as below reported.

\begin{itemize}
    \item[-] The first choice employs the parametric function \cite{2010ApJ...722L.233H}

\begin{equation}\label{TE}
    \eta(z) = 1+\eta_0z,
\end{equation}

that corresponds to a Taylor series around $z\simeq0$. In so doing, we ensure that at $z=0$ $d_A$ and $d_L$ are infinitesimal of the same order around $z\simeq0$.

The immediate drawback with Eq. \eqref{TE} is that data points lie in intervals far from $z\simeq 0$, implying a thorny \emph{convergence problem} \cite{Cattoen:2007id}, plaguing any  Taylor expansions \cite{Aviles:2012ay, Aviles:2014rma, Cattoen:2007sk, Cattoen:2007id}. To face this problem, we below report three more treatments.

\item[-]  The second parametrization consists into a power of $a$, namely a polynomial of the form \cite{2009JCAP...06..012A}

\begin{equation}\label{stepintermedio}
    \eta\equiv (1+z)^{\eta_0},
\end{equation}
where $|\eta_0|\ll1$. Immediately Eq. \eqref{stepintermedio} turns into

\begin{equation}\label{LOG}
    \eta=\exp\{\eta_0\ln(1+z)\}\simeq 1+\eta_0\ln(1+z).
\end{equation}

Since Eq. \eqref{LOG} exhibits a logarithmic correction to the ratio $d_L/d_A$ we hereafter label this logarithmic parametrization with the acronym \emph{LOG}.

The immediate advantage of this paradigm is that, using the logarithm correction, a weaker evolution in terms of $z$ is expected at both $z=0$ and $z\gg1$ regimes, although the convergence problem is not fully-healed.

\item[-] The further treatment accounts for the convergence problem more deeply than the above two parameterizations.

Indeed, we consider the $(1;2)$ Pad\'e series of Eq. \eqref{stepintermedio} that was initially proposed in Ref.  \cite{Gruber:2013wua} with the exact intend of healing the convergence problem in cosmographic expansions.

Subsequently, the different structure of the underlying series exhibits the advantage to converge as $z$ increases since it depends on the two orders, at numerator and denominator, reducing the convergence itself \cite{Gruber:2013wua, Capozziello:2020ctn, Aviles:2014rma}. Precisely, we consider\footnote{The use of a Pad\'e series was already employed in Ref. \cite{Jesus:2024nrl}. In their work the authors investigated three different orders, specifically the $(2;1)$, $(2;2)$ and $(3;2)$ cases concluding, via the bayesian information criterion (BIC)\cite{2007MNRAS.377L..74L}, that lower orders are favored instead of higher ones.}

\begin{equation}\label{P12}
    \eta^{(1;2)}\simeq \frac{1 + {1\over 3} z (2 + \eta_0)}{1 +{2\over3} z (1 - \eta_0) - {1\over 6} z^2 (1 - \eta_0) \eta_0},
\end{equation}

that does not diverge within the interval of data points used in this work. Indeed, any pole of the denominator, $z^\star$, turns out to be $
    z^\star = \frac{2(1-\eta_0)\pm2\sqrt{(1-\eta_0)\left(1+\frac{\eta_0}{2}\right)}}{\eta_0(1-\eta_0)}$ and it is straightforward to check that, for any theoretical value of $\eta_0$, $z^\star$ lies much beyond the lowest and highest values of $z$ adopted throughout the text.

\item[-] The final strategy involves the use of the so-called Chebyshev polynomials $T_n(x)$ which satisfy \cite{chebyshev1854memoires}

\begin{equation}
    T_n(x) = \cos(n\arccos(x)), \quad n\in \mathbb{N}_0.
\end{equation}

Unlike Pad\'e approximants, which are derived from the Taylor expansion of the luminosity distance and exhibit rapidly increasing error bars as the redshift moves away from zero, our approach addresses this limitation, see e.g. Ref. \cite{2018MNRAS.476.3924C}.

Indeed, employing Chebyshev polynomials, we aim to minimize uncertainties in the estimation of cosmographic parameters and, accordingly, on the functional form of $\eta$.

Additionally, we resort to both the Pad\'e and Chebyshev polynomials to investigate even further a possible violation of the CDD relation. In fact, with the exception of Ref. \cite{Jesus:2024nrl}, several efforts have point out that $\eta$ seems not to vary, indicating no CDD violation.

More precisely, we here invoke the recurrence relation generating the polynomials

\begin{equation}\label{chebpol}
    T_{n+1}(x) = 2xT_n(x)-T_{n-1}(x).
\end{equation}
Thus, plugging Eq. \eqref{stepintermedio} in Eq. \eqref{chebpol} and considering $n=1$ we find

\begin{equation}\label{cheb}
    \eta^{(2)} \simeq 2(1+z)^{2\eta_0}-1.
\end{equation}

\end{itemize}

In all the aforementioned four models, we recover the CDD relation in Eq. \eqref{cdd} when $\eta_0=0$.

\section{Methods}\label{cosmicdata}

To test the validity of the CDD relation and obtain constraints on  cosmological parameters, we resort to two methods, the first consisting in a \emph{model-independent approach} while the second, fixing the cosmological model and, \emph{de facto}, appearing as a \emph{model-dependent technique}\footnote{Background cosmology suffers from the issue of degeneration among cosmological models. It is not clear, in fact, which is the best benchmark to describe the late-time acceleration. Consequently, fixing a model \emph{a priori} may lead to degeneracy among scenarios that can furnish analogous results \cite{Efstathiou:1998xx}. To overcome this caveat, one can mix different experimental techniques, using possibly different data, i.e., from early to late-time ones. Alternatively, model-independent treatments can be used to give hints on the right form of the effective underlying models \cite{Nesseris:2010ep, Amendola:2023awr}. Examples are cosmography \cite{Visser:2004bf, Cattoen:2007id, Aviles:2012ay, Dunsby:2015ers, Alfano:2023evg}, Gaussian processes \cite{Shafieloo:2012ht, Zhang:2016tto, Jesus:2023wqk, Seikel:2012uu}, and so on. }.

Moreover, we consider the impact of the DESI data points, considering analyses first with and, then, without their use, associated with other cosmic data sets.

As we report below, we explain in detail the treatments adopted for the analyses.

\subsection{Model independent versus model dependent techniques}

We report our two methodologies and explain their use as follows.

\begin{enumerate}

\item [-] {\bf Model-independent approach}: This method is a useful tool to not assume a cosmological model {\it a priori}. To do that, we resort to the use of the Hubble rate reconstructed via B\'ezier polynomials, first introduced in  Ref. \cite{amati2019addressing} and widely used in the literature \cite{2020A&A...641A.174L, 2021MNRAS.503.4581L, 2023MNRAS.523.4938M, 2023MNRAS.518.2247L, 2024A&A...686A..30A, 2024JCAP...12..055A, 2024arXiv241104878A, 2024arXiv241104901L}. The \emph{ B\'ezier parametric curve} shows a non-linear monotonic growing behavior when stopping at the second order, in agreement with the Hubble rate $H(z)$ trend \cite{2021MNRAS.503.4581L}. In this way, we can write the approximated $H(z)$ as

\begin{equation}\label{bezier}
    H_2(y) = 2g_\alpha\left[\frac{\alpha_0}{2}\left(1-y\right)^2+\alpha_1y\left(1-y\right)+\frac{\alpha_2}{2}y^2\right],
\end{equation}

where $g_\alpha\equiv 100\ \text{km}\cdot \text{s}^{-1}\cdot\text{Mpc}^{-1}$ is a re-scaling factor, $\{ \alpha_0, \alpha_1, \alpha_2 \}$ are B\'ezier coefficients where $\alpha_0\equiv h_0$ with $h_0$, i.e., the reduced Hubble constant, $0 \leq y\equiv z/z_m \leq 1$ with $z_m$ the maximum redshift of the catalog we are going to use.

\item [-] {\bf Model-dependent approach}: In this case we test the CDD relation within background cosmological models, namely the $\Lambda$CDM and $\omega_0\omega_1$CDM models \cite{2020A&A...641A...6P, 2001IJMPD..10..213C, 2003PhRvL..90i1301L}. The general Hubble rate that takes into account the two scenarios is

\begin{equation}\label{modeldip}
    H(z) = H_0\sqrt{\Omega_m\,a^{-3}+\Omega_k\,a^{-2}+\Omega_{de}f(a)},
\end{equation}

with $\Omega_m$, $\Omega_k$, $\Omega_{de} = 1-\Omega_m-\Omega_k$, i.e., the matter, spatial curvature and dark energy magnitudes, respectively. Above,  the dark energy evolution is dictated by  $f(a)$ that is fixed, according to the cosmological model under exam. So, it reduces to unity for the $\Lambda$CDM model while, using $a=(1+z)^{-1}$, to $f(z)=(1+z)^{3(1+\omega_0+\omega_1)}\exp\left(-3\omega_1z(1+z)^{-1}\right)$ for the $\omega_0\omega_1$CDM scenario\footnote{This model has been presented independently by Chevallier and Polarski \cite{2001IJMPD..10..213C} and Linder \cite{2003PhRvL..90i1301L} with the intend of overcoming the convergence issue related to expanding the universe equation of state in function of the redshift. Commonly, the model is also referred to as the CPL parametrization \cite{2007PhRvD..76d1301M, 2007PhRvD..75b3517N} and, with the great exception of DESI findings \cite{2024arXiv240403002D}, was not favored in framing the cosmic expansion history compared with the standard cosmological model, see e.g. Refs. \cite{Sahni:2006pa, 2021PhRvD.104b3520B, 2024JCAP...10..021P}.  }.

\end{enumerate}

\subsection{Cosmic data}

This section is dedicated to illustrate the cosmic data we are going to use in our analyses. In particular, the four data samples used in this work are reported below.

\begin{enumerate}
    \item [-] The updated catalog of OHD encompassing 34 data points, where the values of the Hubble rate at different redshifts $z$ are measured.
    \item [-] The values of the angular-diameter distances inferred by combining observations in the microwave and in the X-ray regimes in 25 galaxy clusters \cite{2005ApJ...625..108D}.
    \item [-] The Pantheon catalog, consisting in 1048 SNeIa from which the magnitudes are calculated \cite{2018ApJ...859..101S}.
    \item [-] The recent BAO catalog as inferred from the \citet{2024arXiv240403002D} consisting in 12 data points divided into 7 redshift bins.
\end{enumerate}

For exhaustive Tables containing the OHD, cluster and BAO data points see e.g. Ref. \cite{2024arXiv241104901L}, while  SNeIa may be found in the dedicated GitHub page, \url{https://github.com/dscolnic/Pantheon}.

\subsubsection{Observational Hubble data}

We employ the recent updated sample of OHD consisting of $N_O = 34$ data points. The Hubble rate measurements are obtained considering at various redshifts $z$ galaxies which evolve passively. Assuming that in these galaxies all stars formed at the same time $\Delta t$, hence the name {\it cosmic chronometers}, and that the redshift is inferred through spectroscopic measurements, the Hubble rate $H(z)$ is approximated by $H(z) \simeq -(1+z)^{-1}(\Delta z/\Delta t)$ \cite{2002ApJ...573...37J, 2018ApJ...868...84M}. Constraints on the quantities of our interest are found by maximizing the following log-likelihood:

\begin{align}
    \ln\mathcal{L}_O = &-\frac{1}{2}\sum^{N_O}_{k=1}\left[\frac{H_k-H(z_k)}{\sigma_{H_k}}\right]^2+&\\&-\frac{1}{2}\sum^{N_O}_{k=1}\ln\left(2\pi\sigma^2_{H_k}\right),\nonumber
\end{align}

where $H_k$ and $\sigma_{H_k}$ are the data points and their errors while $H(z_k)$ changes according to what approach we are considering. Specifically, for the model-independent approach we consider the direct use of Eq. \eqref{bezier}, while for the model-dependent strategy, Eq. \eqref{modeldip}.

\subsubsection{Galaxy cluster data}

The spectral distortion of photons arriving from the cosmic microwave background (CMB) after the recombination epoch due to the interaction via inverse Compton scattering with high-energy electrons in galaxy clusters is known as the \emph{Sunyaev-Zeldovich (SZ) effect} \cite{1970CoASP...2...66S, 1972CoASP...4..173S}.

Combining observations of the CMB photons in the microwave regime together with the X-ray regime, in which the electrons in the intra-cluster are observed, furnishes measurements of the angular-diameter distance, $d_A(z)$, of the clusters. The sample, determined considering the SZ effect, consists in $N_{SZ} = 25$ clusters for which the angular-diameter distances are thus calculated.

However, the combination of the SZ effect, together with measurements in the X-ray to derive the angular-diameter distances, is influenced by the validity of the CDD relation \cite{2004PhRvD..70h3533U}.

The use of such data points is thus essential since they span at higher redshifts and may provide evidence toward CDD. Indeed, as pointed out in Ref.   \cite{2004PhRvD..70h3533U}, to obtain measurements of $d_A(z) = r_c/\theta_c$ with $\theta_c$ the angular size of the cluster, the core radius $r_c$ is needed.

The radius is found considering the ratio between the square of the decrement of the CMB temperature due to the SZ effect, $\Delta T^2_{SZ}$, and the X-ray surface brightness, $S_X$. However, this ratio depends on the validity of the CDD relation as a result of

\begin{equation}
S_X\propto r_c (d_A^2/d_L^2),
\end{equation}
and, so, considering that $\Delta T^2_{SZ}\propto r_c^2$ together with Eq. \eqref{cddviol}, the core radius that it is effectively found, namely $\overline{r}_c$, when the violation of the relation occurs becomes
\begin{equation}
\overline{r}_c = \eta^2r_c.
\end{equation}

Hence, to compute $\eta$, one can either fix the fiducial model by setting constraints on the free parameters or define $\eta$ using one of the four previously discussed parameterizations, as we will do.

Hence, when violating the CDD relation, following the prescriptions provided in Refs. \cite{2011A&A...528L..14H, 2012IJMPD..2150008H}, the corresponding observational angular diameter distance assumes the following form,

\begin{equation}\label{daSZ}
    d^{SZ}_A(z) = d_A(z)\eta^2(z).
\end{equation}

The constraints on the cosmological parameters and on $\eta_0$ are derived by maximizing the following log-likelihood:

\begin{align}
    \ln\mathcal{L}_{SZ} = &-\frac{1}{2}\sum^{N_{SZ}}_{k=1}\left[\frac{d_{A_k}-\eta^2(z_k)d_A(z_k)}{\sigma_{d_{A_k}}}\right]^2+\\&-\frac{1}{2}\sum^{N_{SZ}}_{k=1}\ln\left(2\pi\sigma^2_{d_{A_k}}\right)\nonumber,
\end{align}

where $d_{A_k}$ and $\sigma_{d_{A_k}}$ are the data points and their errors, $\eta^2(z_k)$ assumes the form of one of the four parameterizations described in Eqs. \eqref{TE}-\eqref{LOG}-\eqref{P12} and Eq. \eqref{cheb}, while $d_A(z_k)$ is the angular-diameter distance, say

\begin{equation}\label{da}
    d_A(z) = \frac{c(1+z)^{-1}H_0^{-1}}{\sqrt{|\Omega_k|}}S_k\left[\int^z_0 \sqrt{|\Omega_k|}\frac{H_0}{H(z^\prime)} dz^\prime\right],
\end{equation}

where $H(z)$ is Eq. \eqref{bezier} when we want constraints on the B\'ezier coefficients while it is Eq. \eqref{modeldip} when we investigate the model-dependent scenario. Further, according to the sign of the curvature parameter $\Omega_k$ we have $S_k(x) = \sinh(x)$ if $\Omega_k>0$, $\sin(x)$ if $\Omega_k<0$ and $x$ if $\Omega_k=0$ \cite{1995ApJ...450...14G}.

\subsubsection{Type Ia supernovae}

We employ the Pantheon catalog of SNeIa consisting in $N_S = 1048$ data points \cite{2018ApJ...859..101S}.
Constraints on the cosmological parameters are achieved by maximizing the following log-likelihood \cite{2011ApJS..192....1C}:

\begin{align}
    \ln\mathcal{L}_S = -\frac{1}{2}\left[a+\ln\left(\frac{e}{2\pi}\right)-\frac{b^2}{e}\right],
\end{align}

where $a = \Delta m^T {\bf C_S}^{-1}\Delta m$, $b = \Delta m^T{\bf C_S}^{-1}\mathbf{I}$ and $e=\mathbf{I}^T{\bf C_S}^{-1}\mathbf{I}$. Also, ${\bf C_S}$ is the covariance matrix, $\mathbf{I}$ the identity matrix and $\Delta m = m_k-m(z_k)$ with $m_k$ being the observational magnitudes while $m(z_k)$ is the theoretical one defined as

\begin{equation}
    m(z_k)=25+5\log\left[d_L(z_k)\right].
\end{equation}

Considering Eq. \eqref{cddviol} we can write the previous expression as

\begin{equation}
    m(z_k)=25+5\log\left[\eta(z_k)(1+z_k)^2d_A(z_k)\right],
\end{equation}

where as always, $\eta(z_k)$ takes the form of one of the four parameterizations in Eqs. \eqref{TE}-\eqref{LOG}-\eqref{P12} and Eq. \eqref{cheb} while the angular-diameter distance $d_A(z_k)$ is Eq. \eqref{da}.

\subsubsection{Baryonic acoustic oscillations}

Finally, we resort to the most recent BAO data catalog, namely the DESI sample \cite{2024arXiv240403002D}. The sample consists in $N_{B} = 12$ data points encompassing measurements of the transverse comoving distance $d_M(z)/r_d$, the Hubble rate distance $d_H(z)/r_d$ and a combination of the two labeled averaged distance ratio $d_V(z)/r_d$. For all the distance values $r_d$ is the sound horizon at the drag epoch which we assume to be

\begin{equation}
r_d = (147.09\pm 0.26)\ \text{Mpc},
\end{equation}

from the Planck Collaboration \cite{2020A&A...641A...6P}.

This choice is motivated by the aim of constraining $h_0$ directly rather than $r_dh\equiv H_0r_d(100 \ \text{km}\ \text{s}^{-1}\ \text{Mpc}^{-1})^{-1}$ \cite{2024arXiv240403002D}, thereby reducing the computational complexity. Notably, alternative approaches suggest considering a range of values for $r_d$ that encompasses both the Planck and DESI predictions, see, e.g., Refs. \cite{2024A&A...690A..40L, 2024arXiv240412068C, 2024JCAP...12..055A}.

The distances $d_M(z)$ and $d_V(z)$ can be written in terms of the luminosity distance $d_L(z)$ but, considering that we are assuming a violation of the CDD relation, through Eq. \eqref{cddviol} we have

\begin{subequations}
    \begin{align}
        \frac{d_M(z)}{r_d} &= \frac{\eta(z)d_A(z)(1+z)}{r_d},\label{dm}\\
        \frac{d_H(z)}{r_d}& = \frac{c}{r_dH(z)},\\
        \frac{d_V(z)}{r_d} &= \frac{\left[zd_H(z)\right]^{1/3}\left[d_M(z)\right]^{2/3}}{r_d}.
    \end{align}
\end{subequations}

The constraints on the cosmological parameters are found by maximizing the following log-likelihood:

\begin{align}
    \ln\mathcal{L}_B = &-\frac{1}{2}\sum^{N_B}_{k=1}\left[\frac{X_k-X(z_k)}{\sigma_{X_k}}\right]^2+\\&-\frac{1}{2}\sum^{N_B}_{k=1} \ln\left(2\pi\sigma^2_{X_k}\right),\nonumber
\end{align}

where $X_k$ are the BAO data points with $\sigma_{X_k}$,  the corresponding errors from the catalog while $X(z_k) = \{d_M/r_d,\ d_H/r_d,\ d_V/r_d\}$ varies depending on the cosmological models or B\'ezier interpolation we are going to use.

\section{Numerical findings}\label{results}

To infer constraints on the key cosmological parameters and on $\eta_0$, to argue possible deviations from the CDD relation we perform MCMC analyses using the Metropolis-Hastings algorithm \cite{1953JChPh..21.1087M, 1970Bimka..57...97H}. Our recipe consists in working two fits out, labeled \emph{Analysis 1} and \emph{Analysis 2}. The first fit takes into account the BAO sample from the \citet{2024arXiv240403002D} while, in the second fit, in order to quantify the impact of DESI data, we do not.

\begin{enumerate}
    \item [-] {\bf Analysis 1}: we employ the OHD + SZ + SNeIa + BAO data sets;
    \item [-] {\bf Analysis 2}: we employ the OHD + SZ + SNeIa data sets.
\end{enumerate}

The computations have been performed by considering both the model-dependent and -independent approaches. Our results for the model-independent case for both fits are displayed in Tab. \ref{tab:bfBezier} while Fig. \ref{fig:Bezier} portrays the contour plots. Additionally, Tab. \ref{tab:bfcosmo} shows the results when the flat (non-flat) $\Lambda$CDM and $\omega_0\omega_1$CDM models are assumed together with the corresponding contour plots in Figs. \ref{fig:LCDM}-\ref{fig:CPL} for both \emph{Analysis 1} and \emph{Analysis 2}.

The contour plots have been computed by using \textit{pyGTC}, a Python-based free-available code \cite{Bocquet2016}. Furthermore, in all our Tables, we label the  $\eta(z)$ Taylor expansion with the acronym \emph{TE} while the use of Chebyshev polyonomials is labeled by \emph{T2}.

\begin{table*}
\centering
\setlength{\tabcolsep}{1.2em}
\renewcommand{\arraystretch}{1.1}
\begin{tabular}{lcccc}
\hline
$\alpha_0\equiv h_0$  &  $\alpha_1$ & $\alpha_2$ & $\Omega_k$ & $\eta_0$ \\
\hline\hline
\multicolumn{5}{c}{{\bf Analysis 1}}\\
\hline\hline
\multicolumn{5}{c}{{ TE}}\\
\hline
$0.692^{+0.008(0.016)}_{-0.008(0.016)}$ & $1.113^{+0.025(0.058)}_{-0.028(0.060)}$ & $2.395^{+0.044(0.089)}_{-0.042(0.082)}$ & $0$ & $0.009^{+0.008(0.017)}_{-0.008(0.017)}$\\
$0.690^{+0.010(0.017)}_{-0.006(0.014)}$ & $1.115^{+0.026(0.063)}_{-0.032(0.067)}$ & $2.394^{+0.040(0.093)}_{-0.039(0.081)}$ & $-0.059^{+0.365(0.786)}_{-0.253(0.490)}$ & $0.016^{+0.030(0.060)}_{-0.045(0.084)}$\\
\hline
\multicolumn{5}{c}{{ LOG}}\\
\hline
$0.693^{+0.007(0.015)}_{-0.009(0.017)}$ & $1.112^{+0.031(0.064)}_{-0.028(0.060)}$ & $2.395^{+0.040(0.087)}_{-0.043(0.084)}$ &   $0$ & $0.015^{+0.014(0.028)}_{-0.015(0.030)}$\\
$0.691^{+0.008(0.017)}_{-0.008(0.015)}$ & $1.115^{+0.029(0.064)}_{-0.033(0.069)}$ &
$2.394^{+0.043(0.091)}_{-0.041(0.084)}$ & $-0.002^{+0.194(0.387)}_{-0.139(0.280)}$ & $0.017^{+0.026(0.057)}_{-0.036(0.076)}$\\
\hline
\multicolumn{5}{c}{{ P(1;2)}}\\
\hline
$0.691^{+0.008(0.017)}_{-0.007(0.016)}$ & $1.117^{+0.026(0.058)}_{-0.033(0.064)}$ & $2.388^{+0.049(0.093)}_{-0.036(0.079)}$ & $0$ & $0.015^{+0.012(0.026)}_{-0.014(0.029)}$\\
$0.691^{+0.009(0.016)}_{-0.006(0.015)}$ & $1.117^{+0.027(0.063)}_{-0.034(0.069)}$ & $2.393^{+0.045(0.088)}_{-0.043(0.084)}$ & $-0.008^{+0.209(0.396)}_{-0.141(0.319)}$ & $0.017^{+0.025(0.059)}_{-0.040(0.078)}$\\
\hline
\multicolumn{5}{c}{T2}\\
\hline
$0.691^{+0.009(0.017)}_{-0.007(0.015)}$ & $1.115^{+0.029(0.062)}_{-0.031(0.063)}$ & $2.394^{+0.040(0.085)}_{-0.043(0.082)}$ & $0$ & $0.003^{+0.004(0.008)}_{-0.003(0.007)}$\\
$0.692^{+0.006(0.015)}_{-0.008(0.016)}$ & $1.117^{+0.027(0.061)}_{-0.036(0.069)}$ & $2.390^{+0.044(0.090)}_{-0.036(0.076)}$ & $+0.031^{+0.151(0.342)}_{-0.184(0.335)}$ & $0.002^{+0.007(0.016)}_{-0.008(0.017)}$\\
 \hline\hline
 \multicolumn{5}{c}{{\bf Analysis 2}}\\
\hline\hline
\multicolumn{5}{c}{{ TE}}\\
\hline
$0.705^{+0.021(0.044)}_{-0.021(0.041)}$ & $0.965^{+0.063(0.129)}_{-0.051(0.113)}$ & $2.141^{+0.134(0.297)}_{-0.173(0.337)}$ & $0$ & $-0.023^{+0.027(0.059)}_{-0.029(0.058)}$\\
$0.704^{+0.023(0.046)}_{-0.018(0.039)}$ & $0.909^{+0.096(0.181)}_{-0.054(0.109)}$ & $2.239^{+0.144(0.317)}_{-0.209(0.409)}$ & $+0.922^{+0.460(1.171)}_{-0.957(1.659)}$ & $-0.119^{+0.109(0.190)}_{-0.045(0.100)}$\\
\hline
\multicolumn{5}{c}{{ LOG}}\\
\hline
$0.708^{+0.021(0.045)}_{-0.022(0.043)}$ & $0.946^{+0.074(0.150)}_{-0.063(0.135)}$ & $2.142^{+0.152(0.319)}_{-0.147(0.303)}$ & $0$ & $-0.048^{+0.054(0.105)}_{-0.042(0.088)}$\\
$0.710^{+0.018(0.044)}_{-0.024(0.044)}$ & $0.929^{+0.083(0.179)}_{-0.087(0.175)}$ & $2.173^{+0.201(0.407)}_{-0.152(0.378)}$ & $+0.233^{+0.392(0.880)}_{-0.448(0.800)}$ & $-0.080^{+0.090(0.173)}_{-0.086(0.175)}$\\
\hline
\multicolumn{5}{c}{{ P(1;2)}}\\
\hline
$0.707^{+0.023(0.048)}_{-0.022(0.043)}$ & $0.938^{+0.084(0.161)}_{-0.060(0.135)}$ & $2.157^{+0.128(0.295)}_{-0.170(0.329)}$ & $0$ & $-0.049^{+0.053(0.103)}_{-0.047(0.096)}$\\
$0.706^{+0.023(0.049)}_{-0.019(0.042)}$ & $0.934^{+0.081(0.173)}_{-0.082(0.183)}$ & $2.159^{+0.223(0.426)}_{-0.144(0.349)}$ & $+0.128^{+0.453(0.923)}_{-0.340(0.731)}$ & $-0.070^{+0.063(0.163)}_{-0.086(0.178)}$\\
\hline
\multicolumn{5}{c}{T2}\\
\hline
$0.704^{+0.026(0.050)}_{-0.017(0.039)}$ & $0.957^{+0.068(0.139)}_{-0.080(0.152)}$ & $2.118^{+0.177(0.344)}_{-0.125(0.272)}$ & $0$ & $-0.010^{+0.011(0.023)}_{-0.013(0.026)}$\\
$0.708^{+0.021(0.044)}_{-0.022(0.044)}$ & $0.925^{+0.094(0.182)}_{-0.084(0.180)}$ & $2.223^{+0.162(0.350)}_{-0.190(0.415)}$ & $+0.130^{+0.456(0.897)}_{-0.340(0.744)}$ & $-0.017^{+0.016(0.039)}_{-0.022(0.045)}$\\
 \hline
\end{tabular}
\caption{Best-fit B\'ezier coefficients and $\eta_0$ in both a flat and non-flat scenario with attached errors at $1$-$\sigma$ ($2$-$\sigma$).}
\label{tab:bfBezier}
\end{table*}

\subsection{Analysis 1}

We here analyze the outcomes from the best-fit values of  B\'ezier coefficients, $\alpha_i$, for the model-independent analysis and the best-fit cosmological parameters for the model-dependent analysis.

In addition, the $\eta_0$ best-fit values for both cases are thus found by maximizing the total log-likelihood,

\begin{equation}
    \ln\mathcal{L} = \ln\mathcal{L}_O+\ln\mathcal{L}_{SZ}+\ln\mathcal{L}_S+\ln\mathcal{L}_B.
\end{equation}

Accordingly, we split our discussion about the outcomes inferred from our computations by considering first the model-independent treatment and then the model-dependent scenario.

\begin{enumerate}
    \item [-] {\bf Model-independent approach}.

   The outcomes of our MCMC analysis show that for the flat case the first B\'ezier coefficient $\alpha_0$, which corresponds to the reduced Hubble constant $h_0$, agrees only at $2$-$\sigma$ with the value inferred from the \citet{2020A&A...641A...6P}, i.e., $h_0=0.674\pm 0.005$, while it is not in agreement with the one from Ref. \cite{Riess:2021jrx}, i.e., $h_0 = 0.730\pm 0.010$.

   When we consider the non-flat case, there is no agreement between our inferred $h_0$ values and the Planck $h_0 = 0.636^{+0.021}_{-0.023}$  \cite{2020A&A...641A...6P}.

    Concerning the curvature parameter $\Omega_k$, we compare it with $\Omega_k=-0.011^{+0.013}_{-0.012}$ from the \citet{2020A&A...641A...6P} finding an agreement
    at $1$-$\sigma$ for all four parameterizations.

    Last but not least, we find rough compatibility with  $\eta_0=0$ at $1$-$\sigma$ \emph{only} for the LOG and Chebyshev's parameterizations, within the context of spatially  flat universe. Conversely, in the non-flat case $\eta_0$ agrees at $1$-$\sigma$ with $\eta_0=0$ for all four parameterizations.

    \item [-] {\bf Model-dependent approach}.

    In this case, our results point out that the  $h_0$ values, obtained within our computations, in the flat case, tend to agree at $2$-$\sigma$ \emph{only} with Planck using the $\Lambda$CDM and $\omega_0\omega_1$CDM models with Planck. Further, for both the scenarios, none of our results agree with $h_0$ found by Riess.

    The matter density $\Omega_m$ is in line, at $1$-$\sigma$, with the value from the Ref. \cite{2020A&A...641A...6P}, i.e., $\Omega_m=0.315\pm 0.007$, for what concerns both the models. Moreover, in the flat case, the quantities $\omega_0$ and $\omega_1$ agree at $2$-$\sigma$ \emph{only} with the expectations of the concordance paradigm, namely $\omega_0 = -1$ and $\omega_1 = 0$.

    In turn, considering the non-flat scenario for both the concordance and $\omega_0\omega_1$CDM models we inferred that none of the values of $h_0$ for both cases are in agreement with Planck when $\Omega_k\neq 0$. Moreover, the matter density parameter $\Omega_m$ is compatible with $\Omega_m=0.348^{+0.013}_{-0.014}$ from the \citet{2020A&A...641A...6P} at $1$-$\sigma$ for both cases. Then, the curvature parameter is compatible at $1$-$\sigma$ with the expectation Planck for both the $\Lambda$CDM and $\omega_0\omega_1$CDM scenarios.

    Finally, in the non-flat scheme $\omega_0$ agrees at $1$-$\sigma$ with $\omega_0=-1$ while $\omega_1$ is only compatible at $2$-$\sigma$ with zero.

    Concerning $\eta_0$, for the flat case it agrees at $1$-$\sigma$ with $0$ only for the $\Lambda$CDM model and for the Chebyshev parametrization in the $\omega_0\omega_1$CDM scenario. Nevertheless, Chebyshev polynomials have been mainly employed  to heal systematic uncertainties. Consequently, our results suggest that the observed deviations from $\eta_0=0$, found in some previous studies on the duality relation, may be instead driven by error bars. Last but not least, in the non-flat case, the agreement lies on $1$-$\sigma$ for both the underlying frameworks.

\end{enumerate}

\subsection{Analysis 2}

The best-fit values of the B\'ezier coefficients $\alpha_i$ for the model-independent analysis, the best-fit cosmological parameters for the model-dependent analysis and the best-fit values of $\eta_0$ for both cases are found by maximizing the total log-likelihood

\begin{equation}
    \ln\mathcal{L} = \ln\mathcal{L}_O+\ln\mathcal{L}_{SZ}+\ln\mathcal{L}_S.
\end{equation}

As done for the \emph{Analysis 1}, we discuss the model-independent case and, then, the model-dependent scenario.

\begin{itemize}
    \item [-] {\bf Model-independent approach}.

    Excluding from our analysis the BAO data points from the \citet{2024arXiv240403002D} we find that our $h_0$ is compatible at $1$-$\sigma$ with the value from \citet{Riess:2021jrx} while the agreement with Planck is only at $2$-$\sigma$.

    When $\Omega_k\neq0$, there is no compatibility with  $h_0$ from Planck. Regarding the curvature parameter $\Omega_k$, for all four parameterizations, is compatible  with $\Omega_k = -0.011^{+0.013}_{-0.012}$ at $1$-$\sigma$.

    In this analysis, $\eta_0$ is in agreement at $1$-$\sigma$ with $0$ for all parameterizations in the flat scenario. When the curvature is restored, our $\eta_0$ appears consistent with the CDD relation at $1$-$\sigma$ in the LOG parametrization while for the other three cases the agreement is only at $2$-$\sigma$.

    \item [-] {\bf Model-dependent approach}.

    Considering the flat case, our $h_0$ agrees at $1$-$\sigma$ with \citet{Riess:2021jrx} for both the $\Lambda$CDM and $\omega_0\omega_1$CDM models. On the other hand, the agreement with Planck is only at $2$-$\sigma$ in both cases.

    Concerning $\Omega_m$, we find an agreement at $1$-$\sigma$ for both models with $\Omega_m=0.315\pm 0.007$. Regarding $\omega_0$, the values found from our MCMC analysis are in agreement at $1$-$\sigma$ with $\omega_0=-1$ while the consistency with $0$ for $\omega_1$ is only at $2$-$\sigma$.

    Focusing on the non-flat scenario, for both models we find that our $h_0$ does not agree with $h_0=0.636^{+0.021}_{-0.023}$. The matter density $\Omega_m$ agrees at $1$-$\sigma$ with $\Omega_m=0.348^{+0.013}_{-0.014}$ for both models. Focusing on the curvature parameter, it appears in agreement with $\Omega_k=-0.011^{+0.013}_{-0.012}$ from Planck at $1$-$\sigma$ for both the frameworks considered in this work.

    Afterwards, $\omega_0$ is consistent at $1$-$\sigma$ with the prediction from the concordance paradigm while $\omega_1$ is only consistent at $2$-$\sigma$ with being zero.

   Finally, our $\eta_0$ agrees at $1$-$\sigma$ with $0$ in the flat (non-flat) concordance paradigm.  Considering the flat $\omega_0\omega_1$CDM scenario, we find that $\eta_0$ is consistent with $1$-$\sigma$ for the Taylor and Chebyshev's parameterizations while for the LOG and Pad\'e parameterizations the agreement is only at $2$-$\sigma$ confidence level. When the curvature is accounted for in the $\omega_0\omega_1$CDM case, our values of $\eta_0$ only agree at $2$-$\sigma$ confidence level for all the parameterizations employed in our analysis.
\end{itemize}

\onecolumngrid
\setlength{\tabcolsep}{0.8em}
\renewcommand{\arraystretch}{1.2}

\begin{longtable}{lccccc}
\caption{Best-fit cosmological parameters of the flat (non-flat) $\Lambda$CDM and $\omega_0\omega_1$CDM models and $\eta_0$ with attached errors at $1$-$\sigma$ ($2$-$\sigma$).}
\label{tab:bfcosmo}\\
\hline
$h_0$ & $\Omega_m$ & $\Omega_k$ & $\omega_0$ & $\omega_1$ & $\eta_0$\\
\endfirsthead
\caption{Continued.}\\
\hline
$h_0$ & $\Omega_m$ & $\Omega_k$ & $\omega_0$ & $\omega_1$ & $\eta_0$\\
\hline
\endhead
\endfoot
\hline

\multicolumn{6}{c}{$\Lambda$CDM}\\
\hline\hline
\multicolumn{6}{c}{{\bf Analysis 1}}\\
\hline\hline
\multicolumn{6}{c}{{TE}}\\
\hline
$0.694^{+0.007(0.015)}_{-0.008(0.016)}$ & $0.299^{+0.013(0.027)}_{-0.012(0.024)}$ & $0$ & $-1$ & $0$ & $0.007^{+0.008(0.015)}_{-0.009(0.016)}$\\
$0.694^{+0.007(0.015)}_{-0.009(0.017)}$ & $0.290^{+0.027(0.055)}_{-0.025(0.051)}$ & $+0.023^{+0.076(0.144)}_{-0.066(0.139)}$ & $-1$ & $0$ & $0.005^{+0.009(0.018)}_{-0.008(0.017)}$\\
\hline
\multicolumn{6}{c}{{LOG}}\\
\hline
$0.694^{+0.007(0.015)}_{-0.008(0.015)}$ & $0.298^{+0.012(0.025)}_{-0.012(0.024)}$ & $0$ & $-1$ & $0$ & $0.009^{+0.012(0.025)}_{-0.012(0.025)}$\\
$0.692^{+0.008(0.016)}_{-0.008(0.016)}$ & $0.291^{+0.022(0.047)}_{-0.028(0.053)}$ & $+0.023^{+0.079(0.151)}_{-0.055(0.124)}$ & $-1$ & $0$ & $0.006^{+0.013(0.027)}_{-0.013(0.025)}$\\
\hline
\multicolumn{6}{c}{{P(1;2)}}\\
\hline
$0.694^{+0.008(0.015)}_{-0.007(0.015)}$ & $0.298^{+0.013(0.027)}_{-0.012(0.024)}$ & $0$ & $-1$ & $0$ & $0.008^{+0.013(0.025)}_{-0.011(0.024)}$\\
$0.692^{+0.008(0.017)}_{-0.007(0.016)}$ & $0.283^{+0.029(0.055)}_{-0.020(0.044)}$ & $+0.047^{+0.053(0.123)}_{-0.083(0.148)}$ & $-1$ & $0$ & $0.003^{+0.015(0.029)}_{-0.010(0.023)}$\\
 \hline
\multicolumn{6}{c}{T2}\\
\hline
$0.693^{+0.008(0.016)}_{-0.007(0.015)}$ &  $0.299^{+0.013(0.026)}_{-0.013(0.024)}$ & $0$ & $-1$ & $0$ & $0.002^{+0.003(0.006)}_{-0.003(0.006)}$\\
$0.692^{+0.008(0.016)}_{-0.007(0.015)}$ & $0.288^{+0.024(0.049)}_{-0.024(0.048)}$ & $+0.033^{+0.064(0.139)}_{-0.063(0.131)}$ & $-1$ & $0$ & $0.001^{+0.003(0.007)}_{-0.003(0.006)}$\\
\hline\hline
 \multicolumn{6}{c}{{\bf Analysis 2}}\\
\hline\hline
\multicolumn{6}{c}{{TE}}\\
\hline
$0.702^{+0.021(0.043)}_{-0.020(0.040)}$ & $0.277^{+0.043(0.090)}_{-0.038(0.076)}$ & $0$ & $-1$ & $0$ & $-0.017^{+0.029(0.058)}_{-0.026(0.053)}$\\
$0.704^{+0.020(0.043)}_{-0.020(0.041)}$ & $0.307^{+0.085(0.170)}_{-0.084(0.159)}$ & $-0.075^{+0.170(0.346)}_{-0.146(0.311)}$ & $-1$ & $0$ & $-0.013^{+0.028(0.059)}_{-0.028(0.055)}$\\
\hline
\multicolumn{6}{c}{{LOG}}\\
\hline
$0.701^{+0.021(0.042)}_{-0.021(0.040)}$ & $0.280^{+0.043(0.090)}_{-0.039(0.074)}$ & $0$ & $-1$ & $0$ & $-0.020^{+0.040(0.081)}_{-0.037(0.075)}$\\
$0.706^{+0.020(0.045)}_{-0.022(0.042)}$ & $0.311^{+0.076(0.150)}_{-0.064(0.141)}$ & $-0.088^{+0.148(0.325)}_{-0.150(0.298)}$ & $-1$ & $0$ & $-0.021^{+0.037(0.082)}_{-0.037(0.074)}$\\
\hline
\multicolumn{6}{c}{{P(1;2)}}\\
\hline
$0.701^{+0.020(0.042)}_{-0.020(0.040)}$ & $0.279^{+0.043(0.088)}_{-0.038(0.076)}$ & $0$ & $-1$ & $0$ & $-0.021^{+0.040(0.080)}_{-0.038(0.077)}$\\
$0.708^{+0.019(0.042)}_{-0.024(0.044)}$ & $0.313^{+0.072(0.153)}_{-0.072(0.144)}$ & $-0.100^{+0.171(0.342)}_{-0.127(0.290)}$ & $-1$ & $0$ & $-0.024^{+0.041(0.082)}_{-0.034(0.074)}$\\
\hline
\multicolumn{6}{c}{T2}\\
\hline
$0.701^{+0.021(0.041)}_{-0.020(0.040)}$ &  $0.282^{+0.042(0.089)}_{-0.041(0.078)}$ & $0$ & $-1$ & $0$ & $-0.005^{+0.009(0.020)}_{-0.010(0.019)}$\\
$0.704^{+0.023(0.047)}_{-0.019(0.041)}$ &  $0.319^{+0.068(0.147)}_{-0.077(0.149)}$ & $-0.095^{+0.168(0.341)}_{-0.139(0.300)}$ & $-1$ & $0$ & $-0.005^{+0.009(0.019)}_{-0.010(0.020)}$\\
\hline
 \multicolumn{6}{c}{$\omega_0\omega_1$CDM}\\
 \hline\hline
\multicolumn{6}{c}{{\bf Analysis 1}}\\
\hline\hline
\multicolumn{6}{c}{{TE}}\\
\hline
$0.689^{+0.009(0.018)}_{-0.006(0.016)}$ & $0.321^{+0.012(0.030)}_{-0.025(0.093)}$ & $0$ & $-0.872^{+0.095(0.229)}_{-0.113(0.207)}$ & $-0.961^{+0.921(1.848)}_{-0.588(1.400)}$ & $0.008^{+0.010(0.018)}_{-0.007(0.015)}$\\
$0.691^{+0.008(0.016)}_{-0.008(0.016)}$ & $0.342^{+0.048(0.126)}_{-0.075(0.187)}$ & $-0.104^{+0.270(0.444)}_{-0.191(0.597)}$ & $-0.822^{+0.134(0.269)}_{-0.201(0.548)}$ & $-0.625^{+0.409(1.914)}_{-0.705(1.889)}$ & $0.023^{+0.027(0.080)}_{-0.036(0.052)}$\\
\hline
\multicolumn{6}{c}{{LOG}}\\
\hline
$0.690^{+0.008(0.017)}_{-0.007(0.016)}$ & $0.317^{+0.014(0.033)}_{-0.020(0.065)}$ & $0$ & $-0.872^{+0.114(0.247)}_{-0.105(0.204)}$ & $-0.813^{+0.634(1.439)}_{-0.689(1.514)}$ & $0.013^{+0.016(0.033)}_{-0.010(0.027)}$\\
$0.690^{+0.006(0.015)}_{-0.004(0.014)}$ & $0.333^{+0.005(0.091)}_{-0.072(0.126)}$ & $-0.051^{+0.241(0.414)}_{-0.060(0.258)}$ & $-0.788^{+0.009(0.203)}_{-0.246(0.590)}$ & $-0.980^{+0.611(1.649)}_{-0.788(3.374)}$ & $0.028^{+0.010(0.048)}_{-0.051(0.087)}$\\
\hline
\multicolumn{6}{c}{{P(1;2)}}\\
\hline
$0.690^{+0.008(0.017)}_{-0.006(0.016)}$ & $0.315^{+0.018(0.034)}_{-0.016(0.055)}$ & $0$ & $-0.864^{+0.107(0.244)}_{-0.099(0.215)}$ & $-0.760^{+0.492(1.370)}_{-0.723(1.714)}$ & $0.017^{+0.012(0.026)}_{-0.016(0.032)}$\\
$0.692^{+0.006(0.014)}_{-0.012(0.017)}$ & $0.335^{+0.062(0.112)}_{-0.078(0.224)}$ & $-0.082^{+0.302(0.461)}_{-0.207(0.379)}$ & $-0.829^{+0.153(0.281)}_{-0.236(0.668)}$ & $-0.657^{+0.232(1.984)}_{-1.512(4.280)}$ & $0.027^{+0.045(0.080)}_{-0.048(0.095)}$\\
  \hline
\multicolumn{6}{c}{T2}\\
\hline
$0.692^{+0.005(0.015)}_{-0.009(0.018)}$ &  $0.315^{+0.018(0.034)}_{-0.017(0.070)}$ & $0$ & $-0.853^{+0.109(0.225)}_{-0.116(0.233)}$ & $-0.896^{+0.719(1.608)}_{-0.643(1.478)}$ & $0.004^{+0.003(0.007)}_{-0.004(0.008)}$\\
$0.692^{+0.003(0.014)}_{-0.009(0.016)}$ &  $0.300^{+0.066(0.133)}_{-0.048(0.104)}$ & $+0.060^{+0.144(0.324)}_{-0.205(0.477)}$ & $-0.945^{+0.227(0.386)}_{-0.164(0.463)}$ & $-0.802^{+0.609(1.372)}_{-0.767(3.310)}$ & $0.001^{+0.001(0.025)}_{-0.007(0.016)}$\\
\hline\hline
 \multicolumn{6}{c}{{\bf Analysis 2}}\\
\hline\hline
\multicolumn{6}{c}{{TE}}\\
\hline
$0.705^{+0.014(0.038)}_{-0.025(0.045)}$ & $0.349^{+0.058(0.110)}_{-0.041(0.147)}$ & $0$ & $-0.894^{+0.212(0.478)}_{-0.195(0.371)}$ & $-2.937^{+2.081(3.976)}_{-2.274(5.191)}$ & $-0.036^{+0.038(0.068)}_{-0.015(0.047)}$\\
$0.699^{+0.016(0.038)}_{-0.018(0.041)}$ & $0.212^{+0.217(0.469)}_{-0.107(0.187)}$ & $+0.345^{+0.208(0.472)}_{-0.472(1.305)}$ & $-1.101^{+0.519(0.797)}_{-0.240(0.679)}$ & $-4.935^{+2.850(7.013)}_{-5.953(9.377)}$ & $-0.059^{+0.053(0.155)}_{-0.025(0.065)}$\\
\hline
\multicolumn{6}{c}{{LOG}}\\
\hline
$0.705^{+0.018(0.043)}_{-0.025(0.046)}$ & $0.363^{+0.040(0.092)}_{-0.057(0.140)}$ & $0$ & $-0.949^{+0.227(0.510)}_{-0.189(0.398)}$ & $-3.287^{+2.363(4.477)}_{-1.933(5.580)}$ & $-0.050^{+0.042(0.096)}_{-0.037(0.087)}$\\
$0.704^{+0.019(0.045)}_{-0.025(0.046)}$ & $0.263^{+0.176(0.291)}_{-0.106(0.201)}$ & $+0.245^{+0.194(0.364)}_{-0.455(0.689)}$ & $-0.994^{+0.316(0.791)}_{-0.347(0.899)}$ & $-5.453^{+3.471(6.340)}_{-6.125(9.483)}$ & $-0.072^{+0.065(0.151)}_{-0.057(0.110)}$\\
\hline
\multicolumn{6}{c}{{P(1;2)}}\\
\hline
$0.697^{+0.028(0.054)}_{-0.015(0.036)}$ & $0.371^{+0.036(0.085)}_{-0.056(0.150)}$ & $0$ & $-0.937^{+0.230(0.509)}_{-0.171(0.392)}$ & $-2.955^{+2.005(4.029)}_{-2.472(5.514)}$ & $-0.036^{+0.034(0.082)}_{-0.063(0.105)}$\\
$0.706^{+0.012(0.037)}_{-0.022(0.047)}$ & $0.241^{+0.143(0.337)}_{-0.068(0.179)}$ & $+0.267^{+0.164(0.332)}_{-0.349(0.753)}$ & $-1.027^{+0.297(0.768)}_{-0.271(0.693)}$ & $-4.934^{+3.475(5.916)}_{-4.439(9.962)}$ & $-0.079^{+0.039(0.133)}_{-0.042(0.104)}$\\
 \hline
 \multicolumn{6}{c}{T2}\\
\hline
$0.702^{+0.020(0.045)}_{-0.021(0.044)}$ &  $0.363^{+0.043(0.093)}_{-0.057(0.159)}$ & $0$ & $-0.961^{+0.233(0.537)}_{-0.168(0.368)}$ & $-2.928^{+2.134(4.092)}_{-2.541(5.926)}$ & $-0.012^{+0.012(0.025)}_{-0.011(0.022)}$\\
$0.704^{+0.014(0.044)}_{-0.021(0.045)}$ &  $0.271^{+0.123(0.266)}_{-0.099(0.263)}$ & $+0.214^{+0.232(0.407)}_{-0.287(0.675)}$ & $-1.069^{+0.343(0.901)}_{-0.245(0.724)}$ & $-4.831^{+3.063(7.141)}_{-6.494(10.13)}$ & $-0.022^{+0.015(0.037)}_{-0.011(0.024)}$\\

\hline
\end{longtable}

\twocolumngrid

\section{Model selection criteria}\label{mselec}

Model selection criteria are statistical techniques aiming to identify the preferred cosmological model.

Their necessity arises from the degeneracy between different cosmological models at the background level. Therefore, to identify the statistically favored approach, it is possible to simultaneously compare the chi-squared function with the amount of data required to derive constraints for various models.

In our case, it is also a useful test to further check the validity of the CDD relation and thus, the robustness of the concordance paradigm.

In this work we consider the AIC and the DIC \cite{akaike1998information, spiegelhalter2002bayesian,  2006PhRvD..74b3503K}

\begin{subequations}\label{modelsel}
    \begin{align}
        &\text{AIC}\equiv 2\left(n-\ln\mathcal{L}_m\right),\\
        &\text{DIC}\equiv 2\left(n_{eff}-2\ln\mathcal{L}_m\right),
    \end{align}
\end{subequations}

where $\ln\mathcal{L}_m$ is the maximum log-likelihood inferred from the MCMC analysis in each treatment, $n$ are the number of parameters we are considering varying according to which scenario we are analyzing while $n_{eff} = 2\ln\mathcal{L}_m+\langle-2\ln\mathcal{L}\rangle$, with $\langle-2\ln\mathcal{L}\rangle$ representing the average over the posterior distribution\footnote{Remarkably, it is worth noticing that we do not use the Bayesian information criterion (BIC), defined as $\text{BIC}\equiv -2\ln\mathcal{L}_{m}+2n\ln(N)$,  where $N$ represents the number of data points of the adopted samples. Our choice is motivated by the fact that, unlike DIC, the BIC statistic depends on the total number of the free parameters, \emph{de facto} penalizing the final outcomes as prompted in Ref.  \cite{2007MNRAS.377L..74L}.}.

To derive the statistically favored cosmological model, we check out the differences between each value from Eqs. \eqref{modelsel} labeled as $\text{Y}_i=\{\text{AIC}, \text{DIC}\}$ and the lowest value $\text{Y}_0$, associated with the fiducial model,

\begin{equation}
    \Delta\text{Y} = \text{Y}_i-\text{Y}_0.
\end{equation}

The results inferred from using these criteria for the model-independent approach are displayed in Tab. \ref{tab:modelselmodelind} while the ones for the model-dependent case are shown in Tab. \ref{tab:modelselmodeldip} and compared with the usual intervals for AIC and DIC \cite{2021PhRvD.104b3520B},

\begin{itemize}
    \item [-] $\Delta\text{AIC(DIC)}\in [0, 2]$ hints that the model is weakly favored even if it still questions which scenario is the best;
    \item [-] $\Delta\text{AIC(DIC)}\in (2, 6]$ hints that the model is slightly disfavored;
    \item [-] $\Delta\text{AIC(DIC)}> 6$ hints that the model should be dismissed.
\end{itemize}

The selection criteria for the model-independent methodology suggest that for all parameterizations of $\eta(z)$ and in both analyses there is a preference towards a spatially flat universe even if a slight curvature is not fully-excluded. Accordingly, for the model-dependent case, for all the $\eta$ parameterizations,  the favored cosmological model is the flat $\Lambda$CDM scenario even for \emph{Analysis 1}, when the BAO catalog from the \citet{2024arXiv240403002D} is accounted for. Also in this case, a slight curvature is not excluded while the $\omega_0\omega_1$CDM is the least favored scenario.

\begin{table*}
    \centering
    \setlength{\tabcolsep}{2.3em}
    \renewcommand{\arraystretch}{1.1}
    \begin{tabular}{lccccc}\\
    \hline
    & $\ln \mathcal L_m$ & \text{AIC} & DIC & $\Delta$AIC & $\Delta$DIC\\
    \hline\hline
    \multicolumn{6}{c}{{\bf Analysis 1}}\\
    \hline\hline
    \multicolumn{6}{c}{{TE}}\\
    \hline
      B\'ezier ($\Omega_k=0$) & $-843.66$ & $1695$ & $1705$ & $0$ & $0$\\
     B\'ezier ($\Omega_k\neq 0$) & $-843.70$ & $1697$ & $1707$ & $2$ & $2$\\
     \hline
     \multicolumn{6}{c}{LOG}\\
     \hline
      B\'ezier ($\Omega_k=0$) & $-843.68$ & $1695$ & $1706$ & $0$ & $0$\\
     B\'ezier ($\Omega_k\neq 0$) & $-843.70$ & $1697$ & $1708$ & $2$ & $2$\\
     \hline
     \multicolumn{6}{c}{P(1;2)}\\
     \hline
      B\'ezier ($\Omega_k=0$) & $-843.67$ & $1695$ & $1705$ & $0$ & $0$\\
     B\'ezier ($\Omega_k\neq 0$) & $-843.68$ & $1697$ & $1709$ & $2$ & $4$\\
     \hline
     \multicolumn{6}{c}{T2}\\
     \hline
      B\'ezier ($\Omega_k=0$) & $-843.68$ & $1695$ & $1706$ & $0$ & $0$\\
     B\'ezier ($\Omega_k\neq 0$) & $-843.68$ & $1697$ & $1710$ & $2$ & $5$\\
     \hline\hline
     \multicolumn{6}{c}{{\bf Analysis 2}}\\
     \hline\hline
     \multicolumn{6}{c}{TE}\\
     \hline
     B\'ezier ($\Omega_k=0$) & $-836.61$ & $1681$ & $1682$ & $0$ & $0$\\
     B\'ezier ($\Omega_k\neq 0$) & $-836.29$ & $1682$ & $1683$ & $1$ & $1$\\
     \hline
     \multicolumn{6}{c}{LOG}\\
     \hline
      B\'ezier ($\Omega_k=0$) & $-836.54$ & $1681$ & $1682$ & $0$ & $0$\\
     B\'ezier ($\Omega_k\neq 0$) & $-836.50$ & $1683$ & $1684$ & $2$ & $2$\\
     \hline
     \multicolumn{6}{c}{P(1;2)}\\
     \hline
     B\'ezier ($\Omega_k=0$) & $-836.54$ & $1681$ & $1683$ & $0$ & $0$\\
     B\'ezier ($\Omega_k\neq 0$) & $-836.45$ & $1683$ & $1684$ & $2$ & $2$\\
     \hline
      \multicolumn{6}{c}{T2}\\
     \hline
     B\'ezier ($\Omega_k=0$) & $-836.55$ & $1682$ & $1682$ & $0$ & $0$\\
     B\'ezier ($\Omega_k\neq 0$) & $-836.49$ & $1683$ & $1684$ & $2$ & $2$\\
     \hline
    \end{tabular}
    \caption{Comparison between the flat (non-flat) B\'ezier parametrization. The upper panel shows the privileged cosmological model for \emph{Analysis 1} while the lower panel shows the privileged cosmological model for \emph{Analysis 2} for all four parametrization of $\eta(z)$.}
\label{tab:modelselmodelind}
\end{table*}

\begin{table*}
    \centering
    \setlength{\tabcolsep}{2.5em}
    \renewcommand{\arraystretch}{1.1}
    \begin{tabular}{lccccc}\\
    \hline
    & $\ln \mathcal L_m$ & \text{AIC} & DIC & $\Delta$AIC & $\Delta$DIC\\
    \hline\hline
    \multicolumn{6}{c}{{\bf Analysis 1}}\\
    \hline\hline
    \multicolumn{6}{c}{{TE}}\\
    \hline
     $\Lambda$CDM ($\Omega_k=0$) & $-844.20$ & $1694$ & $1697$ & $0$ & $0$\\
     $\Lambda$CDM ($\Omega_k\neq 0$) & $-844.16$ & $1696$ & $1700$ & $2$ & $3$\\
     $\omega_0\omega_1$CDM ($\Omega_k=0$) & $-843.65$ & $1697$ & $1701$ & $3$ & $4$\\
     $\omega_0\omega_1$CDM ($\Omega_k\neq 0$) & $-843.62$ & $1699$ & $1704$ & $5$ & $7$\\
     \hline
     \multicolumn{6}{c}{LOG}\\
     \hline
     $\Lambda$CDM ($\Omega_k=0$) & $-844.31$ & $1696$ & $1697$ & $0$ & $0$\\
     $\Lambda$CDM ($\Omega_k\neq 0$) & $-844.21$ & $1696$ & $1700$ & $2$ & $2$\\
     $\omega_0\omega_1$CDM ($\Omega_k=0$) & $-843.61$ & $1697$ & $1700$ & $3$ & $3$\\
     $\omega_0\omega_1$CDM ($\Omega_k\neq 0$) & $-843.69$ & $1699$ & $1709$ & $5$ & $12$\\
     \hline
     \multicolumn{6}{c}{P(1;2)}\\
     \hline
     $\Lambda$CDM ($\Omega_k=0$) & $-844.29$ & $1694$ & $1696$ & $0$ & $0$\\
     $\Lambda$CDM ($\Omega_k\neq 0$) & $-844.21$ & $1696$ & $1699$ & $2$ & $2$\\
     $\omega_0\omega_1$CDM ($\Omega_k=0$) & $-843.61$ & $1697$ & $1701$ & $3$ & $4$\\
     $\omega_0\omega_1$CDM ($\Omega_k\neq 0$) & $-843.83$ & $1700$ & $1708$ & $5$ & $10$\\
     \hline
     \multicolumn{6}{c}{T2}\\
     \hline
     $\Lambda$CDM ($\Omega_k=0$) & $-844.31$ & $1695$ & $1699$ & $0$ & $0$\\
     $\Lambda$CDM ($\Omega_k\neq 0$) & $-844.19$ & $1696$ & $1701$ & $2$ & $3$\\
     $\omega_0\omega_1$CDM ($\Omega_k=0$) & $-843.67$ & $1697$ & $1703$ & $3$ & $5$\\
     $\omega_0\omega_1$CDM ($\Omega_k\neq 0$) & $-843.67$ & $1699$ & $1703$ & $5$ & $5$\\
     \hline\hline
     \multicolumn{6}{c}{{\bf Analysis 2}}\\
     \hline\hline
     \multicolumn{6}{c}{TE}\\
     \hline
     $\Lambda$CDM ($\Omega_k=0$) & $-836.88$ & $1680$ & $1680$ & $0$ & $0$\\
     $\Lambda$CDM ($\Omega_k\neq 0$) & $-836.83$ & $1682$ & $1682$ & $2$ & $2$\\
     $\omega_0\omega_1$CDM ($\Omega_k=0$) & $-836.00$ & $1682$ & $1683$ & $2$ & $3$\\
     $\omega_0\omega_1$CDM ($\Omega_k\neq 0$) & $-835.98$ & $1684$ & $1686$ & $4$ & $5$\\
     \hline
     \multicolumn{6}{c}{LOG}\\
     \hline
     $\Lambda$CDM ($\Omega_k=0$) & $-836.92$ & $1680$ & $1680$ & $0$ & $0$\\
     $\Lambda$CDM ($\Omega_k\neq 0$) & $-836.79$ & $1681$ & $1682$ & $2$ & $2$\\
     $\omega_0\omega_1$CDM ($\Omega_k=0$) & $-835.87$ & $1682$ & $1683$ & $2$ & $2$\\
     $\omega_0\omega_1$CDM ($\Omega_k\neq 0$) & $-835.95$ & $1684$ & $1685$ & $4$ & $4$\\
     \hline
     \multicolumn{6}{c}{P(1;2)}\\
     \hline
     $\Lambda$CDM ($\Omega_k=0$) & $-836.91$ & $1680$ & $1680$ & $0$ & $0$\\
     $\Lambda$CDM ($\Omega_k\neq 0$) & $-836.79$ & $1681$ & $1682$ & $2$ & $2$\\
     $\omega_0\omega_1$CDM ($\Omega_k=0$) & $-835.95$ & $1682$ & $1683$ & $2$ & $2$\\
     $\omega_0\omega_1$CDM ($\Omega_k\neq 0$) & $-835.92$ & $1684$ & $1684$ & $4$ & $4$\\
     \hline
     \multicolumn{6}{c}{T2}\\
     \hline
     $\Lambda$CDM ($\Omega_k=0$) & $-836.92$ & $1680$ & $1680$ & $0$ & $0$\\
     $\Lambda$CDM ($\Omega_k\neq 0$) & $-836.80$ & $1681$ & $1682$ & $2$ & $2$\\
     $\omega_0\omega_1$CDM ($\Omega_k=0$) & $-835.95$ & $1682$ & $1682$ & $2$ & $2$\\
     $\omega_0\omega_1$CDM ($\Omega_k\neq 0$) & $-835.82$ & $1684$ & $1684$ & $4$ & $3$\\
     \hline
    \end{tabular}
    \caption{Comparison between the flat (non-flat) $\Lambda$CDM and $\omega_0\omega_1$CDM models. The upper panel shows the privileged cosmological model for \emph{Analysis 1} while the lower panel shows the privileged cosmological model for \emph{Analysis 2} for all four parameterizations of $\eta(z)$.}
\label{tab:modelselmodeldip}
\end{table*}

\section{Conclusions and perspectives}\label{conc}

Immediately after the first data release by the DESI collaboration  \citet{2024arXiv240403002D}, several investigations have shown evidences in favor or against the aforementioned claims that pointed out the need of a slightly evolving dark energy term, different from the standard cosmological paradigm, the $\Lambda$CDM model  \cite{2024arXiv240412068C, 2024JCAP...10..035G, 2024A&A...690A..40L, 2024arXiv241104878A, 2024arXiv241212905C, 2024JCAP...09..062D, 2024JCAP...12..055A, 2024arXiv241104901L, 2024arXiv240408633C, 2024arXiv240917074G, 2024arXiv241201740S}.

In this respect, checking the validity of the CDD duality relation, comparing previous literature with current new data catalogs may open new avenues toward the determination of the correct cosmological model.

The main purpose of this work was exploring the possible evidence for a CDD violation, in view of DESi data, adopting novel forms of expansions associated with the corrections of Etherington's formula.

Particularly, this relation states that the ratio between the angular-diameter distance $d_A$ and the luminosity distance $d_L$ is equal to the inverse square of the scale factor $a(t)$. Hence, if an evident deviation from it is found, robust clues against the standard cosmological model would therefore appear in line with alternative dark energy scenarios.

In this work, we started with the typically-assumed violation,  $d_L(z)/d_A(z)(1+z)^2=\eta(z)$,  where $\eta(z)$ can acquire  different expressions \cite{2017JCAP...09..039H, 2012ApJ...745...98M, 2014JPhCS.484a2035J, Jesus:2024nrl, 2020PhRvD.102f3513D, 2013PhRvD..87j3530E, 2021PhRvD.103j3513A, 2021JCAP...06..052B, 2021MNRAS.502.3500Q, 2012JCAP...06..022H, 2011A&A...528L..14H, 2019PhRvD..99h3523F} and considered \emph{four different parameterizations}, given in particular by:

\begin{itemize}
    \item [i.] A standard Taylor expansion around $z\simeq 0$.
    \item [ii.] A logarithmic parametrization, obtained from a polynomial term and hereafter named LOG model.
    \item [iii.] A rational series, with the aim of reducing the convergence problem, namely the (1;2) Pad\'e expansion.
    \item [iv.] A Chebyshev polynomial, constructed with a cosine basis, of order $n=1$, providing the great advantage of healing the systematics over measurements.
\end{itemize}

Afterwards, we employed two treatments, namely a model-independent (-dependent) method to tackle the CDD violation. Concerning the first stratehy, we adopted the well-established B\'ezier polynomials \cite{amati2019addressing} while for the model-dependent way, we considered as cosmological benchmarks the flat (non-flat) $\Lambda$CDM and $\omega_0\omega_1$CDM scenarios, where the latter appears favored by DESI.

Afterwards, we emphasized the main differences between the model-independent and dependent strategies of analysis, remarking our findings in view of previous results.

To do so, in both cases we worked out a MCMC analysis, employing the Metropolis-Hastings algorithm \cite{1953JChPh..21.1087M, 1970Bimka..57...97H} and  dividing our computations into two main cases, specifically setting the OHD+SNeIa+SZ data sets, plus:

\begin{itemize}
    \item [-] adding the BAO sample from DESI, related to the analysis labeled \emph{Analysis 1},
    \item [-] excluding the  aforementioned sample, within  \emph{Analysis 2}, to check the impact of these novel data points into the computations.
\end{itemize}

Our computations certified that, regarding \emph{Analysis 1}, for both the model-independent (-dependent) scenarios the values of the reduced Hubble constant, $h_0$, inferred from our analysis are in agreement with Planck  at $2$-$\sigma$ only, in the flat scenario. Concerning the cosmological parameters, the matter density, $\Omega_m$, agrees at $1$-$\sigma$ in both the concordance and $\omega_0\omega_1$CDM scenarios, in the flat and non-flat cases.

The curvature parameter is found to agree at $1$-$\sigma$ with Planck in both the model-independent (-dependent) approaches. Then, $\omega_0$ agrees at $1$-$\sigma$ with $\omega_0=-1$ only, in the curved case while $\omega_1$ is only in agreement at $2$-$\sigma$ with zero in both the flat and non-flat scenarios.

On the other hand, for \emph{Analysis 2} we found that in  both the cases, $h_0$ is consistent at $2$-$\sigma$ with Planck. For the cosmological parameters, $\Omega_m$ agrees at $1$-$\sigma$ with both the $\Lambda$CDM and $\omega_0\omega_1$CDM scenarios. Regarding $\Omega_k$, our values turned out to be in agreement at $1$-$\sigma$ for the model-independent (-dependent) cases with the value from Ref. \cite{2020A&A...641A...6P}. Concerning $\omega_0$ and $\omega_1$, the first agrees at $1$-$\sigma$ with $\omega_0=-1$ while the latter agrees with $\omega_1=0$ only at $2$-$\sigma$.

Diving more in the Hubble tension, for \emph{Analysis 1} in the flat case our $h_0$ agrees only at $2$-$\sigma$ with Planck's value and strongly disagrees with the local value, found by Riess. Afterwards, when the BAO sample from DESI has been excluded, it looked compatible at $1$-$\sigma$ with Riess results, instead \cite{Riess:2021jrx} and only at $2$-$\sigma$ with Ref.  \cite{2020A&A...641A...6P}.

Finally, our findings over  $\eta_0$ in the model-independent case agreed at $1$-$\sigma$ with zero, only for the LOG model and Chebyshev polynomials in the flat scenario for \emph{Analysis 1} while the agreement at $1$-$\sigma$ persisted for all the four parameterizations when the BAO points have been excluded. On the other hand, in the model-dependent scenario our $\eta_0$ was in agreement at $1$-$\sigma$ with being vanishing for all the four parameterizations in the flat $\Lambda$CDM model, independently from being computed either in  \emph{Analysis 1} or in \emph{Analysis 2}.

The comparison between the two cosmological models through the use of our model selection criteria favored the \emph{flat concordance paradigm in both our  analyses} reinforcing the previous  constraints obtained on $\eta_0$. Last but not least, adopting the B\'ezier polynomials preferred \emph{a flat universe in both the analyses}. Finally, quite remarkably it is worth noticing that for both the analyzed chances, a slight spatial curvature could not fully-excluded, indicating the need of additional work on refining analyses that include its presence.

Future works will focus on additional data sets, spanning intervals of redshifts that exceed the SNeIa limit, such as intermediate measurements of standard sirens, gamma-ray bursts \cite{Khadka:2021vqa}, and so on.

In addition, in lieu of considering only more $\eta(z)$ parameterizations, we can focus on checking strategies that reduce systematics, in analogy to Chebyshev polynomials, for example considering rational versions of them or more refined cosmographic analyses \cite{Aviles:2016wel,delaCruz-Dombriz:2016bqh}.

\section*{Acknowledgements}

ACA acknowledges the Istituto Nazionale di
Fisica Nucleare (INFN) Sezione di Napoli, Iniziativa Specifica QGSKY. The authors are grateful to Marco Muccino for interesting discussions on the subject of duality problem. This paper is supported by the  Fondazione  ICSC, Spoke 3 Astrophysics and Cosmos Observations. National Recovery and Resilience Plan (Piano Nazionale di Ripresa e Resilienza, PNRR) Project ID $CN00000013$ ``Italian Research Center on  High-Performance Computing, Big Data and Quantum Computing" funded by MUR Missione 4 Componente 2 Investimento 1.4: Potenziamento strutture di ricerca e creazione di ``campioni nazionali di R\&S (M4C2-19 )" - Next Generation EU (NGEU).

\appendix

\section{Contour plots}

This appendix deals with our contours, inferred from the underlying MCMC computations for both the model-independent and -dependent approaches.

\begin{figure*}
\centering
{\hfill
\includegraphics[width=0.48\hsize,clip]{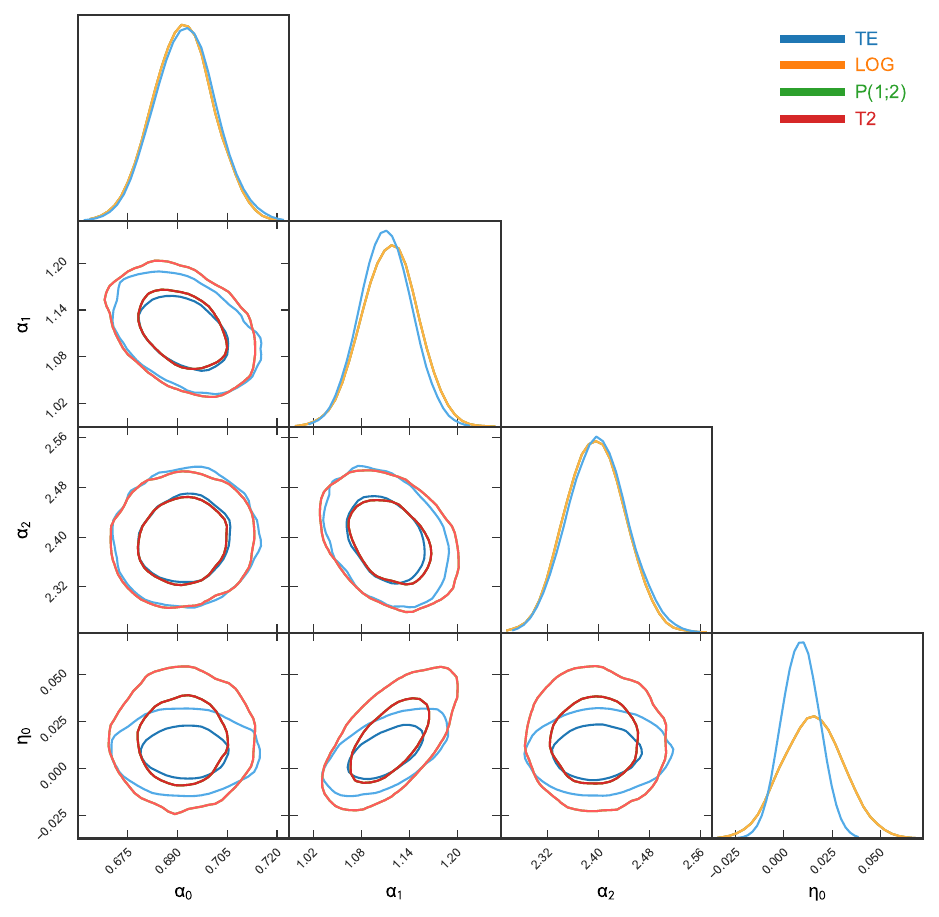}
\hfill
\includegraphics[width=0.48\hsize,clip]{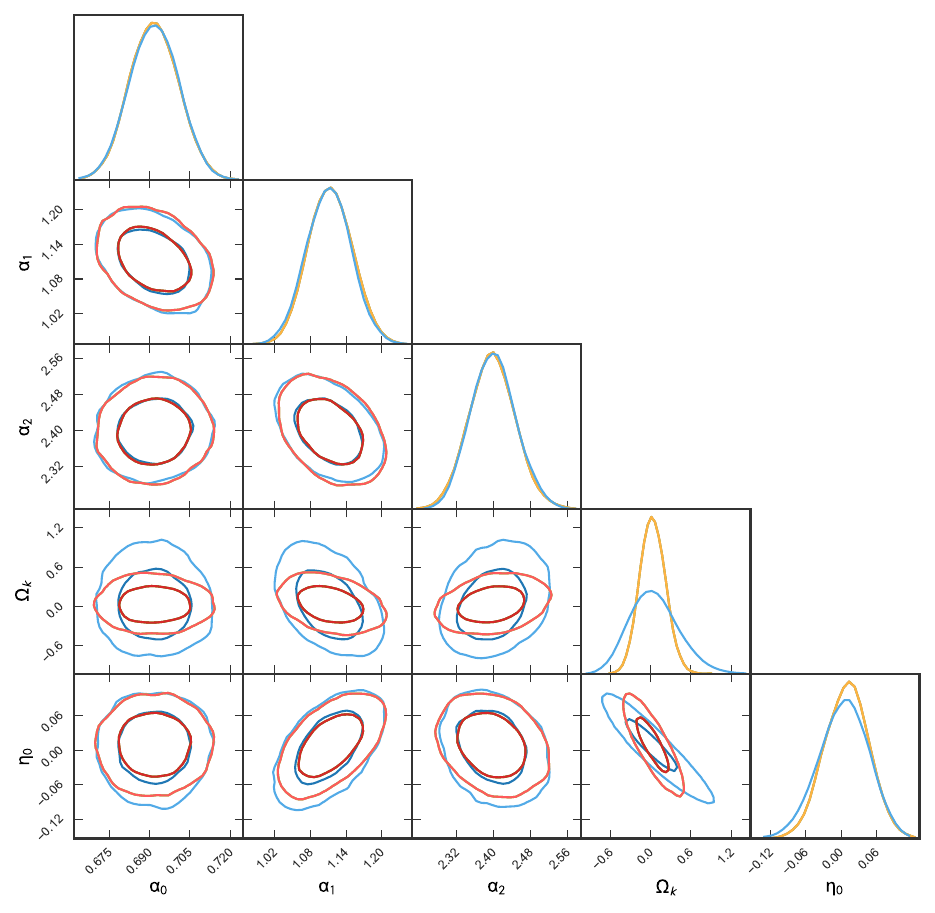}
\hfill}\\
{\hfill
\includegraphics[width=0.48\hsize,clip]{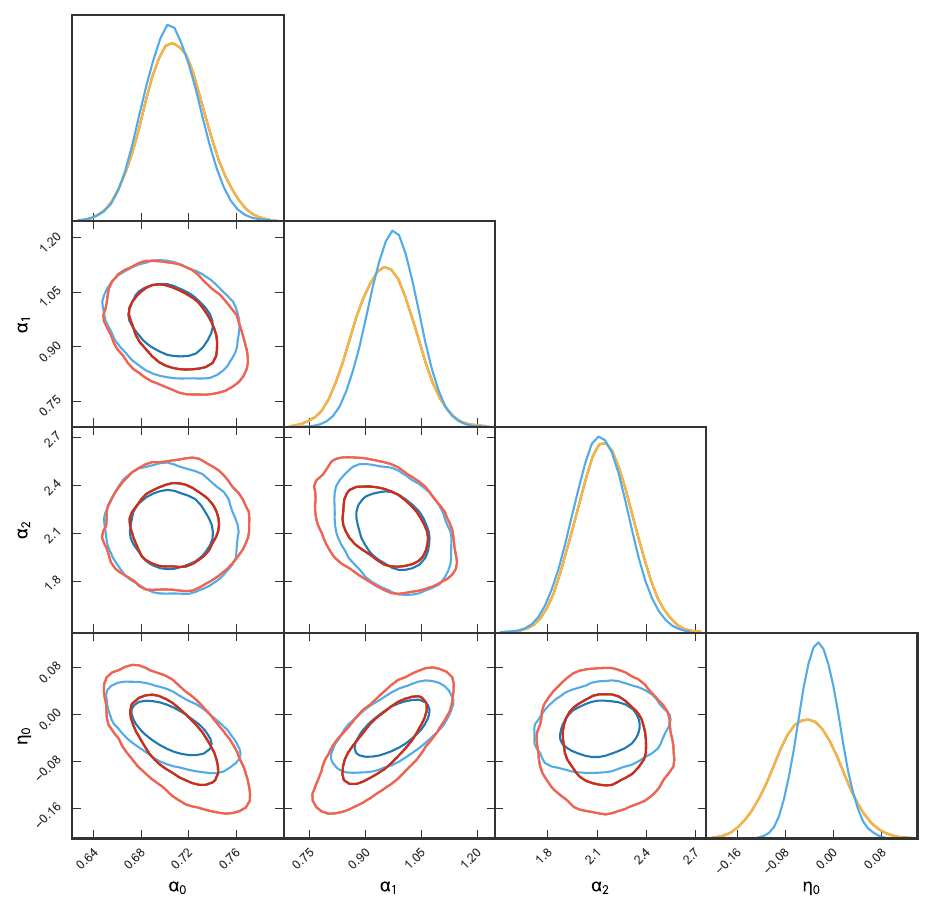}
\hfill
\includegraphics[width=0.48\hsize,clip]{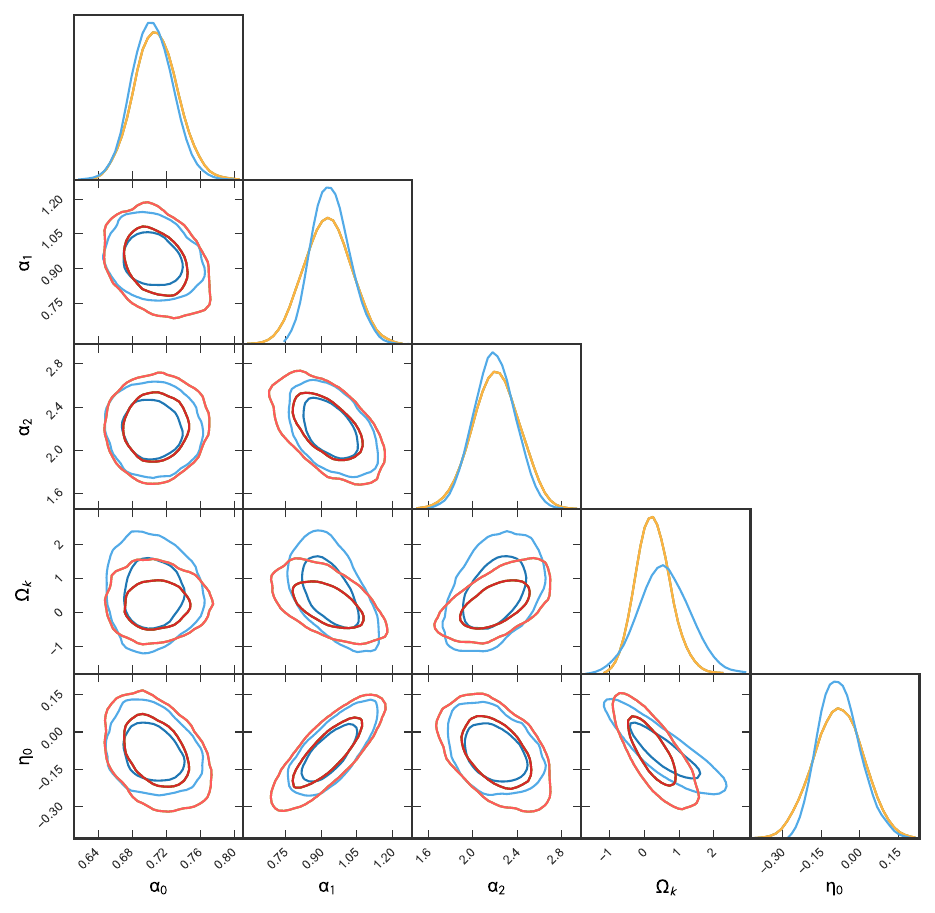}
\hfill}\\
\caption{Contour plots of the best-fit B\'ezier coefficients and $\eta_0$ in both a flat and non-flat scenario. Upper panel shows the contours for \emph{Analysis 1} while the lower panel shows the contours for \emph{Analysis 2}.}
 \label{fig:Bezier}
\end{figure*}

\begin{figure*}
\centering
{\hfill
\includegraphics[width=0.48\hsize,clip]{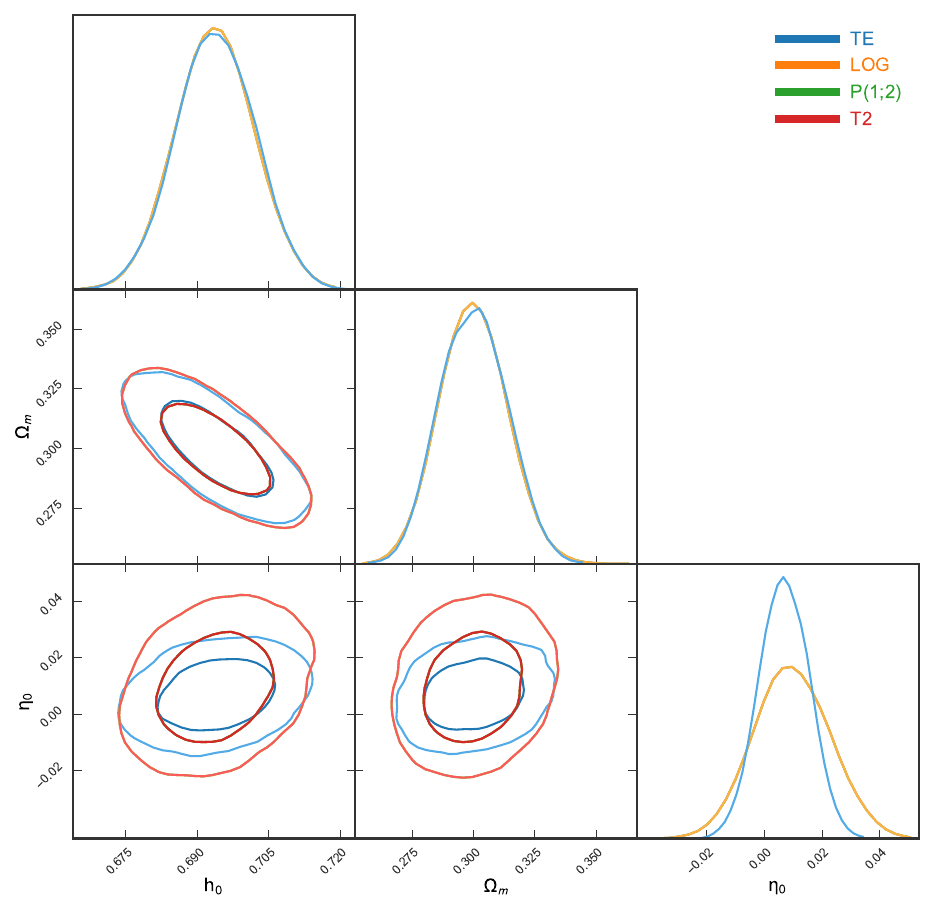}
\hfill
\includegraphics[width=0.48\hsize,clip]{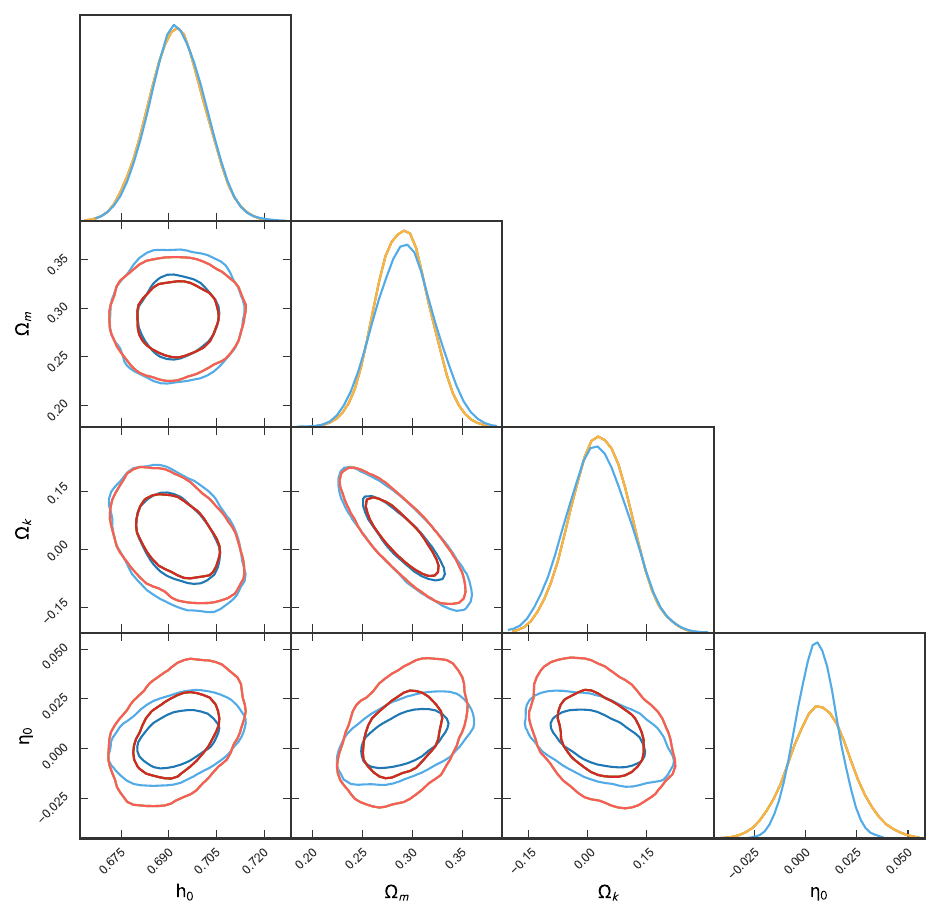}
\hfill}\\
{\hfill
\includegraphics[width=0.48\hsize,clip]{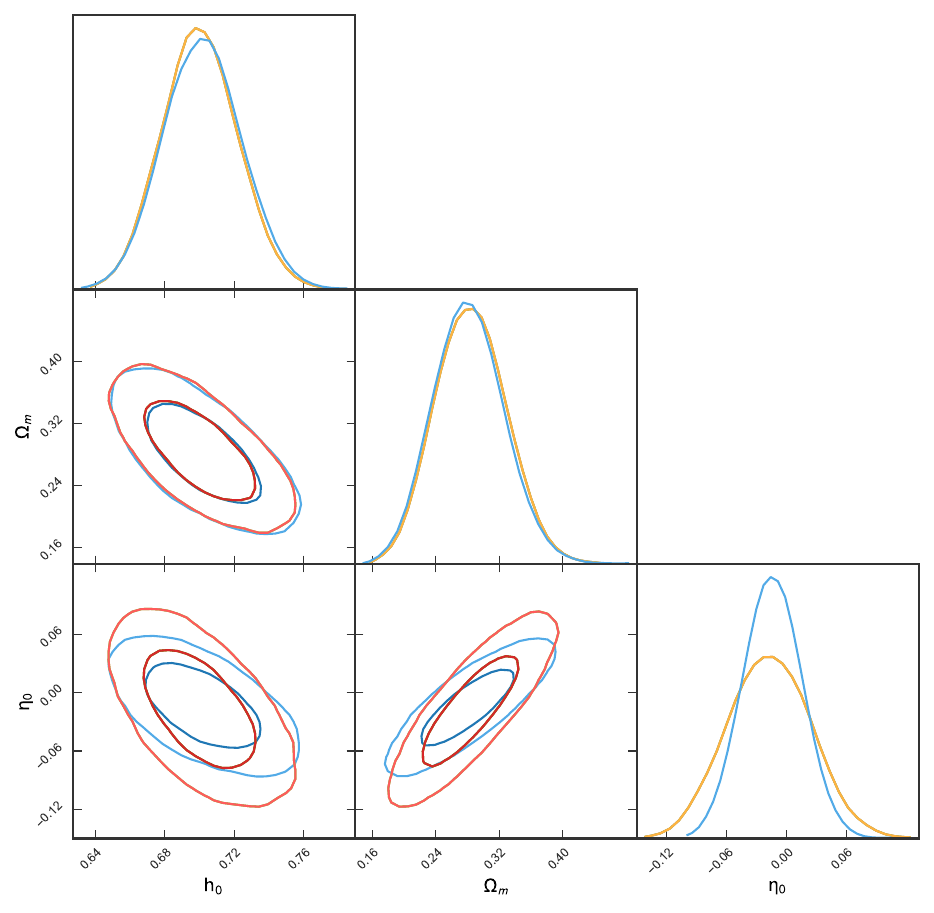}
\hfill
\includegraphics[width=0.48\hsize,clip]{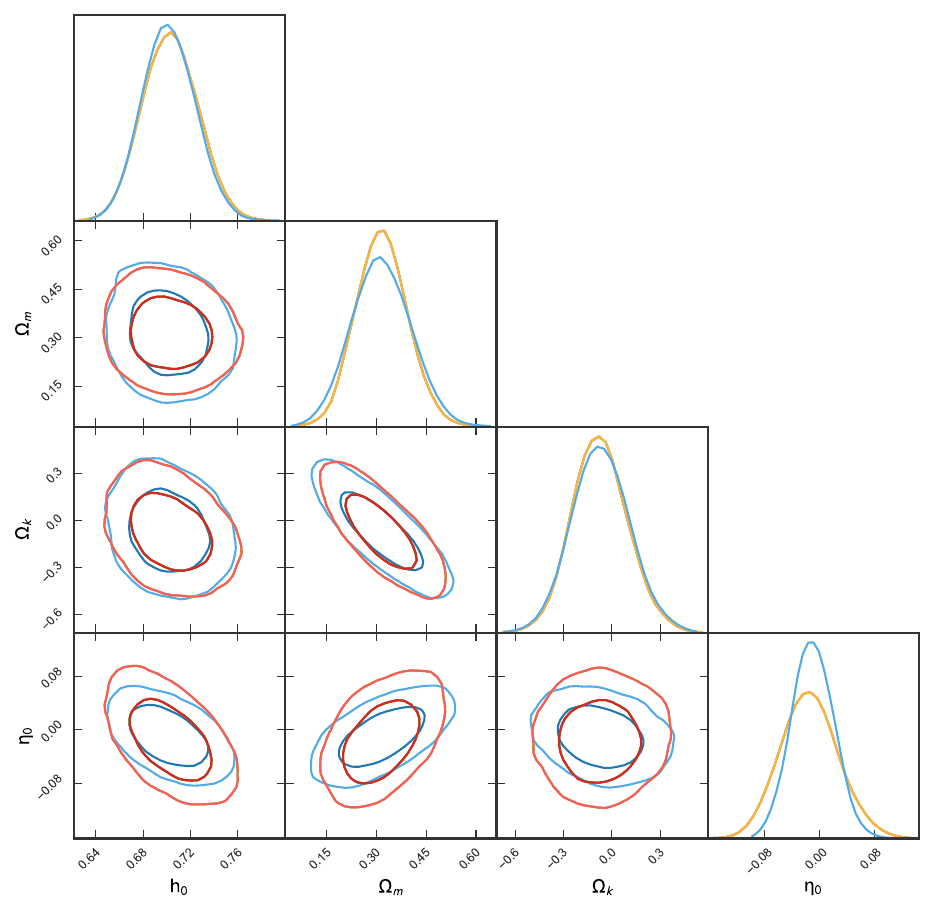}
\hfill}\\
\caption{Contour plots of the best-fit cosmological parameters and $\eta_0$ for the flat (non-flat) $\Lambda$CDM model. Upper panel shows the contours for \emph{Analysis 1} while the lower panel shows the contours for \emph{Analysis 2}.}
 \label{fig:LCDM}
\end{figure*}

\begin{figure*}
\centering
{\hfill
\includegraphics[width=0.48\hsize,clip]{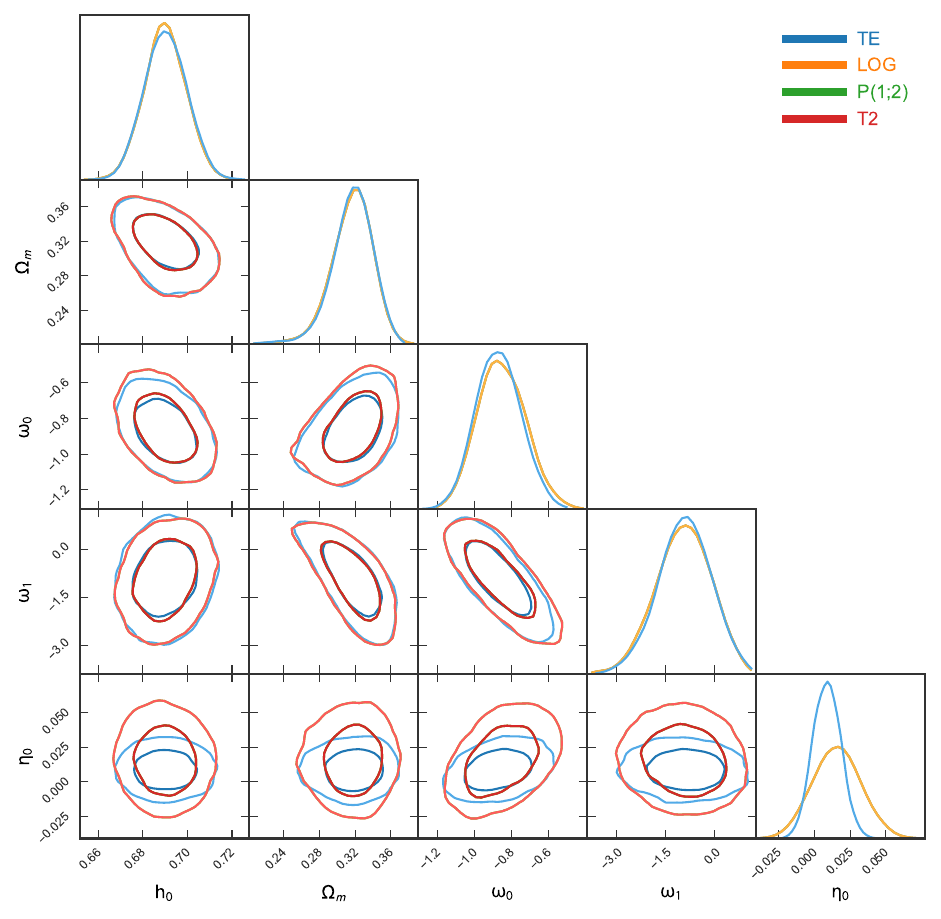}
\hfill
\includegraphics[width=0.48\hsize,clip]{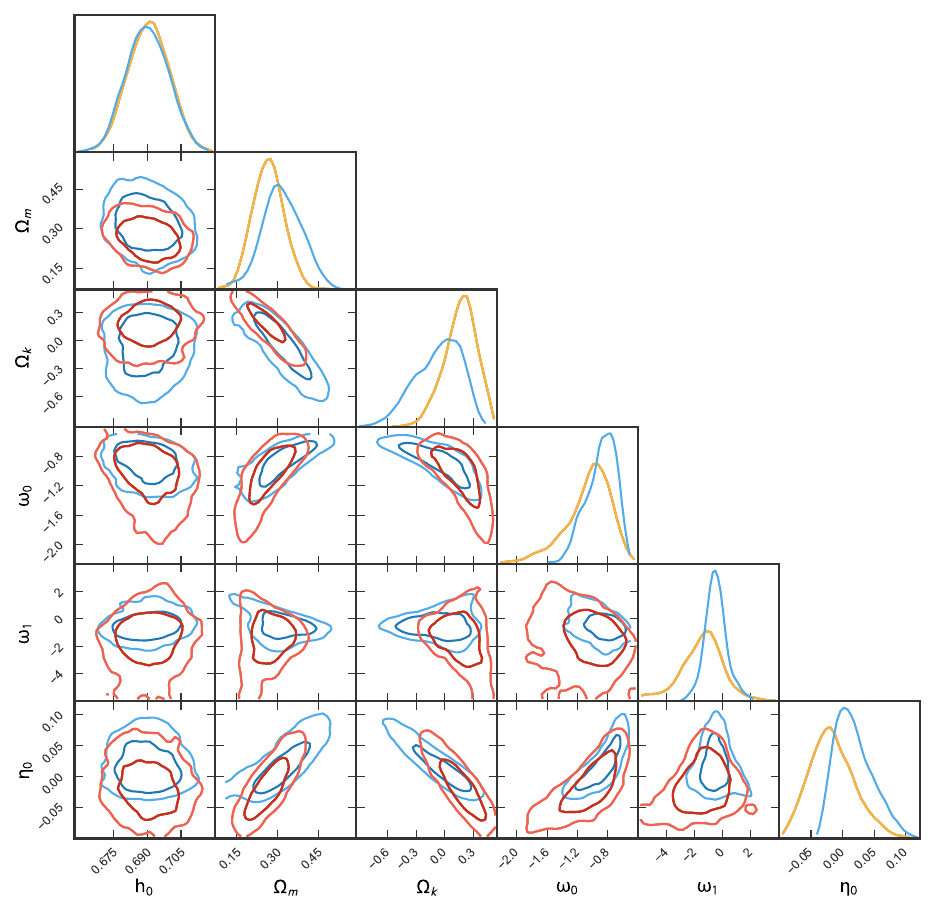}
\hfill}\\
{\hfill
\includegraphics[width=0.48\hsize,clip]{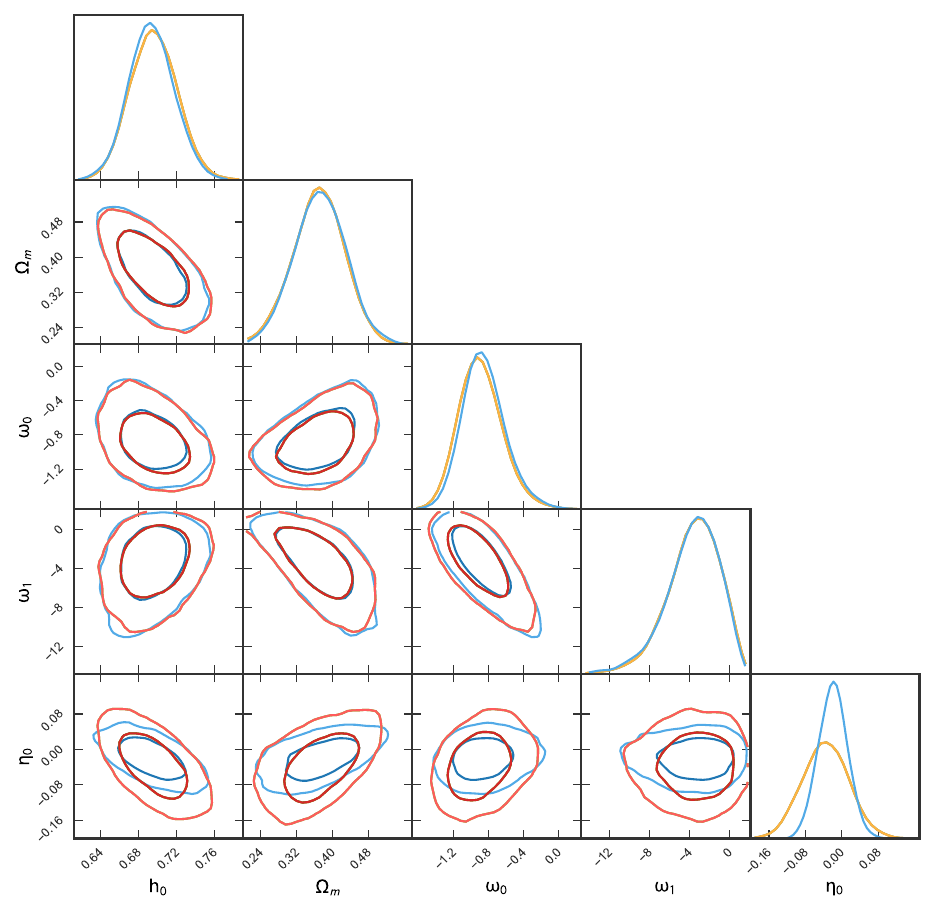}
\hfill
\includegraphics[width=0.48\hsize,clip]{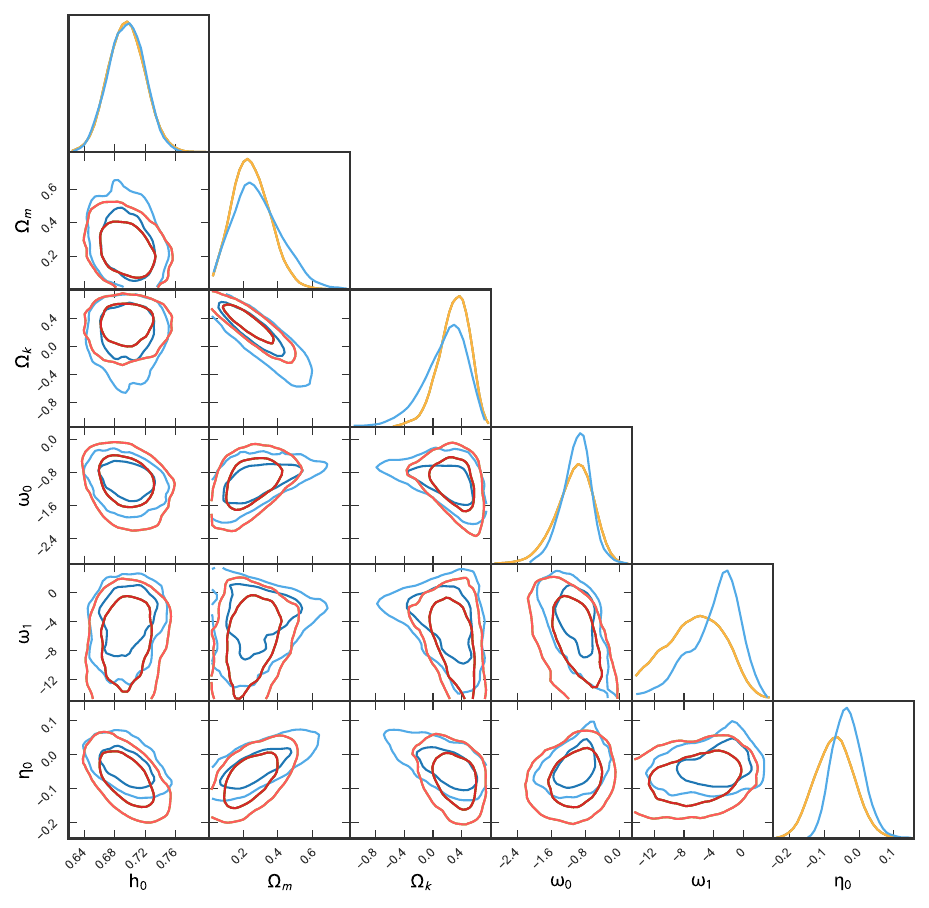}
\hfill}\\
\caption{Contour plots of the best-fit cosmological parameters and $\eta_0$ for the flat (non-flat) $\omega_0\omega_1$CDM model. Upper panel shows the contours for \emph{Analysis 1} while the lower panel shows the contours for \emph{Analysis 2}.}
 \label{fig:CPL}
\end{figure*}

Specifically, Fig. \ref{fig:Bezier} portrays the contour plots for the best-fit B\'ezier coefficients, $\Omega_k$ and $\eta_0$ for both Analyses 1 and 2. Figs. \ref{fig:LCDM}-\ref{fig:CPL} portray the contours for the best-fit cosmological parameters and $\eta_0$ for the flat (non-flat) $\Lambda$CDM and $\omega_0\omega_1$CDM models for both Analyses 1 and 2.

\onecolumngrid

\twocolumngrid

\end{document}